\documentclass[aps,groupedaddress,showpacs,nofootinbib,showkeys,amsfonts,eqsecnum]{revtex4}
\usepackage{bm}

\newcommand{\tr}{{\textrm {tr}}}
\newcommand{\Tr}{{\textrm {Tr}}}
\newcommand{\Det}{{\textrm {Det}}}
\newcommand{\g}{{\textrm {g}}}
\newcommand{\bnu}{{\overline \nu}}
\newcommand{\hnu}{{\widehat \nu}}
\newcommand{\hA}{{\widehat A}}
\newcommand{\htr}{{\widehat\tr}}

\newcommand{\D}{{\widehat D}}
\newcommand{\E}{{\widehat E}}
\newcommand{\F}{{\widehat F}}
\newcommand{\cL}{{\mathcal L}}
\newcommand{\Om}{{\widehat \Omega}}
\newcommand{\ha}{a^{T}} \newcommand{\hb}{b^{T}}
\newcommand{\thruu}[1]{\mathrel{\mathop{#1\!\!\!\!/}}}

\newcommand{\bfx}{\bm{x}}
\newcommand{\bfD}{\bm{D}}
\newcommand{\bfp}{\bm{p}}
\newcommand{\bfA}{\bm{A}}

\newcommand{\half}{{\textstyle\frac{1}{2}}}
\newcommand{\da}{{\dot{a}}}
\newcommand{\db}{{\dot{b}}}
\newcommand{\mod}{{\textrm{mod~}}}

\begin{document}

\title{The thermal heat kernel expansion and the one-loop effective
action of QCD at finite temperature}

\author{E. Meg\'{\i}as}
\email{emegias@ugr.es}

\author{E. \surname{Ruiz Arriola}}
\email{earriola@ugr.es}

\author{L.L. Salcedo}
\email{salcedo@ugr.es}

\affiliation{
Departamento de F\'{\i}sica Moderna, Universidad de Granada, E-18071
Granada, Spain }

\date{\today}

\begin{abstract}
The heat kernel expansion for field theory at finite temperature is
constructed. It is based on the imaginary time formalism and applies
to generic Klein-Gordon operators in flat space-time. Full gauge
invariance is manifest at each order of the expansion and the Polyakov
loop plays an important role at any temperature. The expansion is
explicitly worked out up to operators of dimension six included. The
method is then applied to compute the one loop effective action of QCD
at finite temperature with massless quarks. The calculation is carried
out within the background field method in the $\overline{\text{MS}}$
scheme up to dimension six operators. Further, the action of the
dimensionally reduced effective theory at high temperature is also
computed to the same order. Existing calculations are reproduced and
new results are obtained in the quark sector for which only partial
results existed up to dimension six.
\end{abstract}



\maketitle

\section{Introduction}
\label{sec:1}

The extension of field theory from zero to finite temperature and
density is a natural step undertaken quite early
\cite{Matsubara:1955ws,Kirzhnits:1972ut,Dolan:1974qd,Weinberg:1974hy,%
Polyakov:1978vu,'tHooft:1979uj}. The interest is both at a purely
theoretical level and in the study of concrete physical theories. At
the theoretical level one needs appropriate formulations of the
thermal problem, for which there are several formalisms available
\cite{Landsman:1987uw}, as well as mathematical tools to carry out the
calculations. From the point of view of concrete theories a central
point is the study of the different phases of the model and the nature
of the phase transitions. That study applies not only to condensed
matter theories but also to fundamental ones, such as the electroweak
phase transition, of direct interest in early cosmology and
baryogenesis \cite{Lombardo:2000rs}, and quantum chromodynamics which
displays a variety of phases in addition to the hadronic one
\cite{Gross:1981br,Shaposhnikov:1996th,KorthalsAltes:2003sv,Alford:2002ng}.
Such new phases can presumably be probed at the laboratory in existing
(RHIC) \cite{McLerran:2002jb} and future (ALICE) facilities.
Obviously one expects all these features of QCD at finite temperature
to be fully consistent with manifest gauge invariance. As is well-known Lorentz invariance is manifestly broken due to the privileged
choice of the reference frame at rest with the heat bath, however,
gauge invariance remains an exact symmetry. At zero temperature
preservation of gauge invariance involves mixing of finite orders in
perturbation theory. As will become clear below, compliance with gauge
invariance requires mixing of infinite orders in perturbation theory
at finite temperature.

The purpose of the present work is twofold. The first part (Section
\ref{sec:2}) is devoted to introduce a systematic expansion for the
one-loop effective action of generic gauge theories at finite
temperature in such a way that gauge invariance is manifest at each
order. In the second part this technique is applied to QCD in the high
temperature regime, firstly to compute its one-loop gluon and quark
effective action (Section \ref{sec:3}), and then to derive the
Lagrangian of the dimensionally reduced effective theory (Section
\ref{sec:4}). Further applications can and will be considered in other
cases of interest \cite{Megias:2004prep}.

The effective action, an extension to quantum field theory of the
thermodynamical potentials of statistical mechanics, plays a prominent
theoretical role, being directly related to quantities of physical
interest. To one loop it takes the form $c\Tr\log(K)$, where $K$ is
the differential operator controlling the quadratic quantum
fluctuations above a classical background. Unfortunately, this
quantity is afflicted by mathematical pathologies, such as ultraviolet
divergences or many-valuation (particularly in the fermionic
case). For this reason, it has proved useful to express the effective
action in terms of the diagonal matrix elements of the heat kernel (or
simply the heat kernel, from now on) $\langle x|e^{-\tau K}|x\rangle$,
by means of a proper time representation (see e.g. eq. (\ref{eq:2.17})
below) \cite{Schwinger:1951nm,Dewitt:1975ys}. Unlike the one-loop
effective action, the heat kernel is one-valued and ultraviolet finite
for any positive proper time $\tau$ (we assume that the real part of
$K$ is positive). A further simplifying property is that, after
computing the loop momentum integration implied by taking the diagonal
matrix element, the result is independent of the space-time dimension,
apart from a geometrical factor. In practice the computation of the
heat kernel is through the so called heat kernel expansion. This is an
expansion which classifies the various contributions by their mass
scale dimension, as carried by the background fields and their
derivatives. This is equivalent to an expansion in the powers of the
proper time $\tau$. In this way the heat kernel is written as a sum of
all local operators allowed by the symmetries with certain numerical
coefficients known as Seeley-DeWitt or heat kernel coefficients. The
perturbative and the derivative expansions are two resummations of the
heat kernel expansion. This expansion has been computed to high orders
in flat and curved space-time in manifolds with or without boundary
and in the presence of non-Abelian background fields
\cite{Ball:1989xg,Bel'kov:1996tn,vandeVen:1998pf,Moss:1999wq,%
Fliegner:1998rk,Avramidi:1991je,Gusynin:1989ky}.

In order to apply the heat kernel technique to the computation of the
effective action at finite temperature it is necessary to extend the
heat kernel expansion to the thermal case. This can be done within the
imaginary time formalism, which amounts to a compactification of the
Euclidean time coordinate. The space-time becomes a topological
cylinder. (As usual in this context, we consider only flat space-times
without boundary.)  Now, the heat kernel describes how an initial
Dirac delta function in the space-time manifold spreads out as the
proper time passes, with the Klein-Gordon operator $K$ acting as a
Laplacian operator. As it is known, the standard small $\tau$
asymptotic expansion is insensitive to global properties of the
space-time manifold. This means that the space-time compactification,
and hence the temperature, will not be seen in the strict expansion in
powers of $\tau$. (As a consequence, the ultraviolet sector and hence
the renormalization properties of the theory and the quantum anomalies
are temperature independent, a well-known fact in finite temperature
field theory \cite{Kapusta:1989bk,LeBellac:1996bk}.) Within a path
integral formulation of the propagation in proper time, this
corresponds to an exponential suppression (namely, of order
$e^{-\beta^2/4\tau}$, cf. (\ref{eq:2.5})) of closed paths which wind
around the space-time cylinder. The compactification is made manifest
if instead of counting powers of $\tau$ one classifies the
contributions by their mass dimension. The corresponding thermal
Seeley-DeWitt coefficients will then be powers of $\tau$ but with
exponentially suppressed $\tau$-dependent corrections. As a result of
the compactification, the new expansion will not be Lorentz invariant,
although rotational invariance will be maintained. In addition, we
find coefficients of half-integer order which at zero temperature can
appear only for manifolds with boundary (as distributions with support
on the boundary \cite{McAvity:1991we}). Such half-order terms vanish
in a strict proper time expansion.

Another relevant issue is the preservation of gauge invariance. At
zero temperature the only local gauge covariant quantities available
are the matter fields, the field strength tensor and their covariant
derivatives. However, at finite temperature there is a further gauge
covariant quantity which plays a role, namely, the (untraced) thermal
Wilson line or Polyakov loop. Since temperature effects in the
imaginary time formalism come from the winding around the space-time
cylinder, the Polyakov loop appears naturally in the thermal heat
kernel. Our calculation, anticipated in \cite{Megias:2002vr}, shows
that the thermal heat kernel coefficients at a point $x$ become
functions of the untraced Polyakov loop that starts and ends at
$x$. Although such a dependence is consistent with gauge invariance at
finite temperature, is not required by it either. Nevertheless, there
is a simple argument which shows that the heat kernel expansion cannot
be simply given by a sum of gauge covariant local operators (albeit
with Lorentz symmetry broken down to rotational symmetry). For the
Klein-Gordon operator describing a gas of identical particles free
from any external fields other than a chemical potential (plus a
possible mass term) it is obvious that such chemical potential (which
can be regarded as a constant c-number scalar potential $A_0$) has no
effect through the covariant derivatives, and so it is invisible in
the gauge covariant local operators. However, it is visible in the
Polyakov loop, and it is only in this way that the effective action,
or the grand-canonical potential, and hence the particle density, can
depend on the chemical potential. The dependence of the thermal heat
kernel coefficients on the Polyakov loop was overlooked in previous
calculations \cite{Boschi-Filho:1992ah,Xu:1993qi}, although was made
manifest in particular cases and configurations in
\cite{Actor:2000hc}. Of course, the relevance of the Polyakov loop is
well-known in quarkless QCD at high temperature, where it is the order
parameter signaling the presence of a deconfining phase
\cite{Polyakov:1978vu}. The determination of the effective action of
the Polyakov loop after integration of all others degrees of freedom
has been pursued e.g. in \cite{Pisarski:2000eq}.  Our results imply
that, because the formulas are quite general and should hold for any
gauge group, the Polyakov loop must be accounted for, not only in the
color degrees of freedom and at high temperature, but also in other
cases such as the chiral flavor group with vector and axial-vector
couplings and at any finite temperature \cite{Megias:2004prep}. The
thermal heat kernel expansion is derived in Section \ref{sec:2}.

In Section \ref{sec:3} we apply the previous technique to the
computation of the effective action of QCD at finite temperature to
one-loop. Here we refer to the effective action in the technical sense
of generating function of one-particle irreducible diagrams. For the
quark sector (we consider massless quarks for simplicity) the method
applies directly by taking as Klein-Gordon operator the square of the
Dirac operator and using an integral representation for the fermionic
determinant. In the gluon sector, the fluctuation operator is of the
Klein-Gordon type in the Feynman gauge, and so the technique applies
too, but this time in the adjoint representation of the gauge group
and including the ghost determinant. The calculation is carried out
using the covariant background field method. To treat ultraviolet
divergences dimensional regularization is applied, plus the
$\overline{\text{MS}}$ scheme. We have also made the calculation using
the Pauli-Villars scheme as a check. In this computation the
background gauge fields are not stationary, and this allows to write
expressions which are manifestly invariant under all gauge
transformations (recall that in the time compactified space-time there
are topologically large gauge transformations
\cite{Alvarez-Estrada:1993jm}). The result is expressed using gauge
invariant local operators, including operators of up to dimension 6,
and the Polyakov loop $\Omega(x)$. This is done for arbitrary SU$(N)$
($N$ being the number of colors). For SU(2) and SU(3) the traces on
the color group are worked out, to dimension six for SU(2) and to
dimension four for SU(3). In our expansion the dependence on the
Polyakov loop is treated exactly (we keep all orders in an expansion
in powers of $\log(\Omega)$) but the expansion in covariant
derivatives is truncated. In particular the time covariant derivative
is not kept to all orders and in this we differ from
\cite{Diakonov:2003yy} (who only treat SU(2) for stationary
configurations).

As it is known, the effective action of perturbative QCD at finite
temperature contains infrared divergences due to the massless gluons
in the chromomagnetic sector \cite{Linde:1980ts,Braaten:1995na}. Such
divergences come from stationary quantum fluctuations which are light
even at high temperature, whereas the non stationary modes become
heavy, with an effective mass of the order of the temperature $T$,
from the Matusbara frequency. So, the procedure which has been devised
to avoid the infrared problem is to integrate out the heavy,
non-stationary modes to yield the action of an effective theory for
the stationary modes, i.e. of gluons in three Euclidean dimensions
\cite{Ginsparg:1980ef,Appelquist:1981vg,Nadkarni:1983kb,Landsman:1989be,%
Braaten:1996jr,Kajantie:1996dw,Shaposhnikov:1996th,KorthalsAltes:2003sv}.
In this way one obtains a dimensionally reduced theory ${\cal
L}_{3D}$. (One can go further and integrate out the chromoelectric
gluons which become massive through the Debye mechanism. We do not
consider such further reduction here.)  By construction, ${\cal
L}_{3D}$ reproduces the static Green functions of the four-dimensional
theory ${\cal L}_{4D}$.  Of course, the infrared divergences will
reappear now if this action is used in perturbation theory. However,
living in a lower dimension, ${\cal L}_{3D}$ is better behaved in the
ultraviolet and also more amenable to non perturbative techniques,
such as lattice gauge theory. The parameters of ${\cal L}_{3D}$
(masses, coupling constants) can be computed in standard perturbative
QCD since they are infrared finite, coming from integration of the
heavy non stationary modes, although they are scale dependent due to
the standard ultraviolet divergences of four-dimensional QCD. Section
\ref{sec:4} is devoted to obtaining the action of the reduced
theory. This is easily done from the calculation of the effective
action in Section \ref{sec:3} by removing the static Matsubara mode in
the gluonic loop integrations. This theory inherits the gauge
invariance under stationary gauge transformations of the
four-dimensional theory, but a larger gauge invariance is no longer an
issue since more general gauge transformations would not preserve the
stationarity of the fields. In addition, at high temperature
fluctuations of the Polyakov loop far from unity (or from a center of
the gauge group element in the quarkless case) are suppressed and so
it is natural to expand the action in powers of $A_0$. We obtain the
action up to operators of dimension six included (counting each gluon
field as mass dimension one) and compare with existing calculations to
the same order quoted in the literature
\cite{Landsman:1989be,Chapman:1994vk,Huang:1995cu,Shaposhnikov:1996th,%
Wirstam:2001ka,KorthalsAltes:2003sv}. The relevant scales
$\Lambda^T_{M,E}$ for the running coupling constant in the high
temperature regime are identified and reproduced
\cite{Huang:1995cu}. For the dimension six terms, in the gluon sector
we find agreement with \cite{Chapman:1994vk} if the Polyakov loop is
expanded in perturbation theory and in the quark sector we reproduce
the results of \cite{Wirstam:2001ka} for the particular case
considered there (no chromomagnetic gluons and no more that two
spatial derivatives). We give the general result for SU($N$) and
simpler expressions for the cases of SU(2) and SU(3).

The heat kernel and the QCD parts of the paper may interest different
audiences, the first one being more methodological and the second one
more phenomenological, and to some extent they can be read
independently. The QCD part does not require all the details of the
derivation of the thermal heat kernel expansion but only the final
formulas. In fact, one of the points of this paper is that the thermal
coefficients need not be computed each time for each application but
only once, and then applied in a variety of situations.

\section{The Heat Kernel expansion at finite temperature}
\label{sec:2}

\subsection{The Polyakov loop and the heat kernel}
\label{sec:2.a}

We will consider Klein-Gordon operators of the form
\begin{equation}
K= M(x)-D_\mu^2\,,\quad
D_\mu= \partial_\mu+A_\mu(x)\,.
\end{equation}
$M(x)$ is a scalar field which is a Hermitian matrix in internal space
(gauge group space), the gauge fields $A_\mu(x)$ are antihermitian
matrices. $K$ acts on the particle wave function in $d+1$ Euclidean
dimensions and in the fundamental representation of the gauge
group. At finite temperature in the imaginary time formalism the time
coordinate is compactified to a circle, i.e., the space-time has
topology ${\cal M}_{d+1}=S^1\times {\cal M}_d$. Correspondingly, the
wave functions are periodic in the bosonic case, with period $\beta$
(the inverse temperature) antiperiodic in the fermionic case, and the
external fields $M$, $A_\mu$ are periodic.

In order to obtain the heat kernel $\langle x|e^{-\tau K}|x\rangle$ (a
matrix in internal space) we use the symbols method, extended to
finite temperature in \cite{Salcedo:1998sv,Garcia-Recio:2000gt}: For
an operator $\hat f=f(M,D_\mu)$ constructed out of $M$ and $D_\mu$,
\begin{equation}
\langle x| f(M,D_\mu) |x\rangle = \frac{1}{\beta}\sum_{p_0}
\int\frac{d^dp}{(2\pi)^d} \langle x|f(M,D_\mu+i p_\mu)| 0\rangle \,.
\label{eq:2}
\end{equation}
Here $p_0$ are the Matsubara frequencies, $2\pi n/\beta$ for bosons
and $2\pi (n+\frac{1}{2})/\beta$ for fermions, and the sum extends to
all integers $n$. On the other hand, $|0\rangle$ is the zero momentum
wave function, $\langle x|0\rangle=1$. The matrix valued function
$\langle x|f(M,D_\mu+i p_\mu)| 0\rangle$ is the symbol of $\hat f$. Is
important to note that this wave function is periodic (in fact
constant) and not antiperiodic, even for fermions. The antiperiodicity
of the fermionic wave function is only reflected in the Matsubara
frequencies in this formalism.  Whenever the symbols method is used,
$\partial_\mu$ acts on the periodic external fields. Ultimately
$\partial_\mu$ acts on $|0\rangle$ giving zero (this means in practice
a right-acting derivative operator).

In order to introduce the necessary concepts gradually, and to provide
the rationale for the occurrence of the Polyakov loop in the simplest
case, in what remains of this subsection we will consider the case of
no vector potential, space independent scalar potential and constant
c-number mass term,
\begin{equation}
\bfA(x)=0\,,\quad A_0=A_0(x_0)\,,\quad M(x)= m^2\,,\quad [m^2,~]=0 \,.
\label{eq:3a}
\end{equation}
This choice avoids complications coming from the spatial covariant
derivatives and commutators at this point of the discussion. The
result will be the zeroth order term of an expansion in the number of
commutators $[D_\mu,~]$ and $[M,~]$.

An application of the symbols method yields in this case
\begin{eqnarray}
\langle x| e^{-\tau K}|x\rangle 
&=&
\frac{1}{\beta}\sum_{p_0}
\int\frac{d^dp}{(2\pi)^d} 
 \langle x|e^{-\tau (m^2+\bfp^2-(D_0+i p_0)^2)}| 0\rangle 
\nonumber \\
&=& 
\frac{e^{-\tau m^2}}{(4\pi\tau)^{d/2}}
\frac{1}{\beta}\sum_{p_0}
 \langle x| e^{\tau (D_0+i p_0)^2}| 0 \rangle \,. 
\label{eq:2a}
\end{eqnarray}
(After the replacement $\bfD\to\bfD+\bfp$ dictated by (\ref{eq:2})
$D_i=\partial_i$ can be set to zero due to $|0\rangle$.)

The sum over the Matsubara frequencies implies that the operator
$\frac{1}{\beta}\sum_{p_0} e^{\tau (D_0+i p_0)^2}$ is a periodic
function of $D_0$ with period $2\pi i/\beta$, thus it is actually a
one-valued function of $e^{-\beta D_0}$. This can be made explicit by
using Poisson's summation formula, which yields
\begin{equation}
\frac{1}{\beta}\sum_{p_0} e^{\tau (D_0+i p_0)^2} =
\frac{1}{(4\pi\tau)^{1/2}}
\sum_{k\in \mathbb{Z}} (\pm)^k e^{-k\beta D_0} e^{-k^2\beta^2/4\tau}
\label{eq:2.5}
\end{equation}
($\pm$ for bosons or fermions, respectively). This observation allows
to apply the operator identity \cite{Garcia-Recio:2000gt}
\begin{equation}
 e^{\beta \partial_0}e^{-\beta D_0} = \Omega(x) \,,
\label{eq:1}
\end{equation}
where $\Omega(x)$ is the thermal Wilson line or untraced Polyakov loop
\begin{equation}
\Omega(x)= T\exp\left(-\int_{x_0}^{x_0+\beta}
A_0(x_0^\prime,\bfx)dx_0^\prime\right)
\label{eq:1b}
\end{equation}
($T$ refers to temporal ordering and the definition is given for a
general scalar potential $A_0(x)$.) The Polyakov loop appears here as
the phase difference between gauge covariant and non covariant time
translations around the compactified Euclidean time. Physically, the
Polyakov loop can be interpreted as the propagator of heavy particle
in the gauge field background. The identity (\ref{eq:1}) is trivial
if one chooses a gauge in which $A_0$ is time independent (which
always exists globally) since in such a gauge $\Omega= e^{-\beta
A_0}$, and $D_0$, $A_0$ and $\partial_0$ all commute. The identity
itself is gauge covariant and holds in any gauge
\cite{Garcia-Recio:2000gt}.

The point of using (\ref{eq:1}), is that the translation operator in
Euclidean time, $e^{\beta\partial_0}$, has no other effect than moving
$x_0$ to $x_0+\beta$ and this operation is the identity in the
compactified time,
\begin{equation}
e^{\beta\partial_0}=1\,,
\end{equation}
(even in the fermionic case, recall that after applying the method of
symbols the derivatives act on the external fields and not on the
particle wave functions), so one obtains the remarkable result
\begin{equation}
e^{-\beta D_0} = \Omega(x) \,.
\label{eq:1a}
\end{equation}
That is, whenever the differential operator $D_0$ appears periodically
(with period $2\pi i/\beta$), it can be replaced by the multiplicative
operator (i.e. the ordinary function)
$-\frac{1}{\beta}\log(\Omega(x))$. The many-valuation of the logarithm
is not effective due to the assumed periodic dependence. Another point
to note is that $D_0$ (or any function of it), acts as a gauge
covariant operator on the external fields $F(x_0,\bfx)$, and so
transforms according to the local gauge transformation at the point
$(x_0,\bfx)$. Correspondingly, the Polyakov loop, which is also gauge
covariant, starts at time $x_0$ and not at time zero in (\ref{eq:1b});
this difference would be irrelevant for the traced Polyakov loop, but
not in the present context.

An application of the rule (\ref{eq:1a}), yields in particular,
\begin{equation}
\frac{1}{\beta}\sum_{p_0} e^{\tau (D_0+i p_0)^2} =
\frac{1}{(4\pi\tau)^{1/2}}
\sum_{k\in \mathbb{Z}} (\pm)^k \Omega^k e^{-k^2\beta^2/4\tau} \,.
\label{eq:4}
\end{equation}
More generally, 
\begin{equation}
\sum_{p_0} f(ip_0 + D_0) = \sum_{p_0}f(ip_0-\frac{1}{\beta}\log(\Omega)) \,,
\label{eq:4b}
\end{equation}
provided the sum is absolutely convergent, so that the sum is a periodic
function of $D_0$. Thus 
it will prove useful to introduce the quantity $Q$ defined as
\begin{equation}
Q = ip_0 + D_0 = ip_0-\frac{1}{\beta}\log(\Omega) \,.
\label{eq:3}
\end{equation}
The second equality holds in expressions of the form (\ref{eq:4b}).
(Note that the two definitions of $Q$ are not equivalent in other
contexts, e.g. in $\sum_{p_0}f_1(Q)X f_2(Q)$ unless $[D_0,X]=0$.)

The heat kernel in (\ref{eq:2a}) becomes
\begin{eqnarray}
\langle x| e^{-\tau K}|x\rangle &=&
\frac{1}{(4\pi\tau)^{d/2}}
e^{-\tau m^2}
\frac{1}{\beta}\sum_{p_0}
e^{\tau Q^2}
\label{eq:13a} \\
&=& 
\frac{1}{(4\pi\tau)^{(d+1)/2}}e^{-\tau m^2}
\varphi_0(\Omega)\,.
\label{eq:13}
\end{eqnarray}
In the first equality we have removed the bracket $\langle x|\cdot|
0\rangle$ since for multiplicative operators like $\Omega(x)$, this
brackets just picks up the value of the function at $x$. In the last
equality we have used the definition of the functions
$\varphi_n(\Omega)$ which will appear frequently below
\begin{eqnarray}
\varphi_n(\Omega;\tau/\beta^2) &=&
\left(4\pi\tau\right)^{1/2}\frac{1}{\beta}\sum_{p_0} \tau^{n/2} Q^n
e^{\tau Q^2} \,,\quad
Q = ip_0-\frac{1}{\beta}\log(\Omega) \,.
\label{eq:2.15}
\end{eqnarray}
Note that there is a bosonic and a fermionic version of each such
function, and the two versions are related by the replacement
$\Omega\to -\Omega$. As indicated, these functions depend only of the
combination $\tau/\beta^2$. In the zero temperature limit, the sum
over $p_0$ becomes a Gaussian integral, yielding
\begin{equation}
\varphi_n(\Omega;0) = \left\{
\matrix{(-\frac{1}{2})^{n/2}(n-1)!! & (\text{$n$ even})
\cr 0 & (\text{$n$ odd}) }
\right.
\end{equation}
As can be seen for instance from (\ref{eq:4}), in this limit only the
$k=0$ mode remains, whereas the other modes become exponentially
suppressed, either at low temperature or low proper time $\tau$.

The result in (\ref{eq:13}) is sufficient to derive the grand
canonical potential of a gas of relativistic free particles. For
definiteness we consider the bosonic case \cite{Haber:1981fg}. The
effective action (related to the grand canonical potential through
$W=\beta\Omega_{\text{gc}}$) is obtained as
\begin{equation}
W=\Tr\log(K)= -\Tr\int_0^\infty\frac{d\tau}{\tau} \langle x|e^{-\tau
K}|x\rangle
\label{eq:2.17}
\end{equation}
$K$ includes a chemical potential $A_0=-i\mu$ as unique external
field, a the corresponding Polyakov loop is $\Omega=
\exp(i\beta\mu)$. Using (\ref{eq:13}), subtracting the zero
temperature part (which corresponds to setting $\varphi_0\to 1$) and
carrying out the integrations yields the standard result
\cite{Kapusta:1989bk}
\begin{equation}
W= N\int\frac{d^dxd^dk}{(2\pi)^d}\left[
\log\left(1-e^{-\beta(\omega_k-\mu)}\right)
+\log\left(1-e^{-\beta(\omega_k+\mu)}\right)
\right]\,.
\end{equation}
 $N$ is the number of 
species and $\omega_k=\sqrt{k^2+m^2}$.

In next subsection, after the introduction of more general external
fields, we will consider expansions in the number of spatial covariant
derivatives and mass terms. At zero temperature, the derivative
expansion involves temporal derivatives as well, as demanded by
Lorentz invariance, but such an expansion is more subtle at finite
temperature. The direct method would be to expand in powers of $D_0$
in (\ref{eq:2a}), however, this procedure spoils gauge invariance
(e.g. $D_0|0\rangle= A_0|0\rangle$ is not gauge covariant). As a rule,
giving up the periodic dependence in $D_0$ breaks gauge invariance
\cite{Garcia-Recio:2000gt}. One can try to first fix the gauge so that
$A_0$ is stationary, and then expand in powers of $A_0$. This is
equivalent to expanding in powers of $\log(\Omega)$. By construction
this procedure preserves invariance under infinitesimal (or more
generally, topologically small) gauge transformations, however, it
does not preserve invariance under discrete gauge transformations
(\cite{Garcia-Recio:2000gt,Salcedo:2002pr} and \ref{subsec:3.d}
below). This is because $\log(\Omega)$ is many-valued under such
transformations. An expansion in the number of temporal covariant
derivatives which does not spoil one-valuation nor gauge invariance is
described next.

\subsection{The diagonal thermal heat kernel coefficients}
\label{sec:2.b}

Here we will consider the heat kernel expansion at finite
temperature in the completely general case of non trivial and non
Abelian gauge and mass term fields, $A_\mu(x)$, $M(x)$.

First of all one has to specify the counting of the expansion. At zero
temperature, the expansion is defined as one of $\langle x|e^{-\tau
K}|x\rangle$ in powers of $\tau$ [after extracting the geometrical
factor $(4\pi\tau)^{-(d+1)/2}$]. Each power of $\tau$ is tied to a
local operator constructed with the covariant derivatives $D_\mu$ and
$M(x)$ (cf. (\ref{eq:18}) and (\ref{eq:20})). The heat kernel
$e^{-\tau K}$ is dimensionless by assigning engineering mass
dimensions $-2,+1,+2$ to $\tau$, $D_\mu$ and $M$, respectively. So, at
zero temperature, the expansion in powers of $\tau$ is equivalent to
counting the mass dimension carried by the local operators.

At finite temperature there is a further dimensional quantity,
$\beta$, the two countings are not longer equivalent and one has to
specify the concrete expansion to be used. It is well-known that the
finite temperature corrections are negligible in the ultraviolet
region, so that for instance, the temperature does not modify the
renormalization properties of a quantum field theory
\cite{Kapusta:1989bk,LeBellac:1996bk} and also the quantum anomalies
are not affected \cite{Dolan:1974qd,GomezNicola:1994vq}. The
ultraviolet limit corresponds to the small $\tau$ limit in the heat
kernel. As noted before, and can be seen e.g. in (\ref{eq:4}), the
finite $\beta$ and small $\tau$ corrections are of the order of
$e^{-\beta^2/4\tau}$ or less, and so they are exponentially
suppressed. Of course, the same exponential suppression applies to the
low temperature and finite $\tau$ limit. This implies that a strict
expansion of the heat kernel in powers of $\tau$ will yield precisely
the same asymptotic expansion as at zero temperature. In order to pick
up non trivial finite temperature corrections we arrange our expansion
according to the mass dimension of the local operators. In this
counting we take the Polyakov loop $\Omega$, $D_\mu$ and $M$ as
zeroth, first and second order, respectively. In addition one has to
specify that $\Omega(x)$ is at the left in all terms (equivalently,
one could define a similar expansion with $\Omega$ always at the
right). This is required because the commutator of $\Omega$ with other
quantities generates commutators $[D_0,~]$ which are dimensionful in
our counting. After these specifications the expansion of $\langle
x|e^{-\tau K}|x\rangle$ for a generic gauge group is unique and
well-defined and full gauge invariance is manifest at each order. In
this expansion the terms are ordered by powers of $\tau$ but with
coefficients which depend on $\beta^2/\tau$ and $\Omega$,
\begin{equation}
\langle x|e^{-\tau(M-D_\mu^2)}| x\rangle= 
(4\pi\tau)^{-(d+1)/2}\sum_n
a^T_n(x)\tau^n \,.
\label{eq:18b}
\end{equation}
From the definition it is clear that the zeroth order term for a
general configuration is just
\begin{equation}
a_0^T(x) = \varphi_0(\Omega(x);\tau/\beta^2)
\label{eq:13b}
\end{equation}
already computed in the previous subsection (cf. (\ref{eq:13})). This
is because when the particular case (\ref{eq:3a}) is inserted in the
full expansion all terms of higher order, with one or more $[D_\mu,~]$
or $m^2$, vanish identically.

For subsequent reference we introduce the following notation. The
field strength tensor is defined as $F_{\mu\nu}=[D_\mu,D_\nu]$ and
likewise, the electric field is $E_i=F_{0i}$. In addition, the
notation $\D_\mu$ means the operation $[D_\mu,~]$. Finally we will
use a notation of the type $X_{\mu\nu\alpha}$ to mean
$\D_\mu\D_\nu\D_\alpha X =[D_\mu,[D_\nu,[D_\alpha, X]]]$,
e.g. $M_{00}=\D_0^2M$, $F_{\alpha\mu\nu}=\D_\alpha F_{\mu\nu}$.

The method for expanding a generic function $\langle x| f(M,D_\mu)
|x\rangle$ has been explained in detail in
\cite{Garcia-Recio:2000gt}. We have applied this procedure to compute
the heat kernel coefficients to mass dimension 6. However, for the
heat kernel there is an alternative approach which uses the well-known
Seeley-DeWitt coefficients at zero temperature. This is the method
that we explain in detail here. The idea is as follows. The symbols
method formula (\ref{eq:2}) is applied to the temporal dimension only
\begin{equation}
\langle x|e^{-\tau (M-D_\mu^2)} | x\rangle
= \frac{1}{\beta}\sum_{p_0} 
\langle \bfx |e^{-\tau (M-Q^2-D_i^2)} | \bfx\rangle \,,\quad
Q = ip_0 + D_0
\label{eq:17}
\end{equation}
(The bracket $\langle x_0| \ |0\rangle$, associated to the Hilbert
space over $x_0$, is understood although not written explicitly.) This
implies that we can use the standard zero temperature expansion for
the $d$-dimensional heat kernel with effective Klein-Gordon operator
\begin{equation}
K_0=Y-D_i^2\,, \quad Y=M-Q^2 \,.
\end{equation}
In this context $Y$ is the non Abelian mass term, because, although it
contains temporal derivatives (in $Q$), it does not contain spatial
derivatives and so acts multiplicatively on the spatial Hilbert
space. The standard heat kernel expansion gives then
\begin{equation}
\langle \bfx|e^{-\tau(Y-D_i^2)}| \bfx\rangle= 
(4\pi\tau)^{-d/2}\sum_{n=0}^\infty
a_n(Y,\D_i)\tau^n \,,
\label{eq:18}
\end{equation}
where the coefficients $a_n(Y,\D_i)$ are polynomials of dimension $2n$
made out of $Y$ and $\D_i=[D_i,~]$. To lowest orders
\cite{Ball:1989xg,vandeVen:1998pf}
\begin{eqnarray}
a_0 &=& 1 \,, \nonumber \\
a_1 &=& -Y \,, \nonumber \\
a_2 &=& \frac{1}{2}Y^2-\frac{1}{6}Y_{ii}+\frac{1}{12}F_{ij}^2
\,, \nonumber \\
a_3 &=& -\frac{1}{6}Y^3
+\frac{1}{12}\{Y,Y_{ii}\}
+\frac{1}{12}Y_i^2
-\frac{1}{60}Y_{iijj}
-\frac{1}{60}[F_{iij},Y_j]
\nonumber \\ &&
-\frac{1}{30}\{Y,F_{ij}^2 \}
-\frac{1}{60}F_{ij}YF_{ij}
+\frac{1}{45}F_{ijk}^2
-\frac{1}{30}F_{ij}F_{jk}F_{ki}
+\frac{1}{180}F_{iij}^2
+\frac{1}{60}\{ F_{ij},F_{kkij} \}
 \,.
\label{eq:20}
\end{eqnarray}
(As noted before $Y_{ii}=\D_i^2Y$, $F_{ijk}=\D_i F_{jk}$, etc.)

Eq. (\ref{eq:18}) inserted in (\ref{eq:17}) is of course correct but
not very useful as it stands. For instance, for the zeroth order, the
expansion in (\ref{eq:18}) would be needed to all orders to reproduce
the simple result (\ref{eq:13b}), since $e^{\tau Q^2}$ is not a
polynomial in $Q$. In view of this, we consider instead
\begin{equation}
\langle \bfx|e^{-\tau(M-Q^2-D_i^2)}| \bfx\rangle= 
(4\pi\tau)^{-d/2}\sum_{n=0}^\infty
e^{\tau Q^2} \tilde a_n(Q^2,M,\D_i)\tau^n \,,
\label{eq:18a}
\end{equation}
which introduces a new set of polynomial coefficients $\tilde
a_n(Q^2,M,\D_i)$. By their definition, it is clear that these
coefficients are unchanged if ``$Q^2$'' is everywhere replaced by
``$Q^2+\text{c-number}$''. This implies that in $\tilde a_n$ the
quantity $Q^2$ appears only in the form $[Q^2,~]$. This is an
essential improvement over the original coefficients $a_n$, since each
$[Q^2,\ ]$ will yield at least one $\D_0$, and so higher orders in
$[Q^2,\ ]$ appear only at higher orders in the heat kernel
expansion\footnote{This kind of resummations are standard also at zero
temperature to move e.g. the mass term $e^{-\tau M}$ to the left and
leave only a $[M,~]$ dependence in the coefficients
\cite{Ball:1989xg}.}.

The calculation of the coefficients $\tilde a_n(Q^2,M,\D_i)$ follows
easily from the relation
\begin{equation}
\sum_{n=0}^\infty a_n\tau^n = e^{\tau Q^2} \sum_{n=0}^\infty \tilde
a_n\tau^n \,.
\label{eq:19}
\end{equation}
If one takes the expression on the l.h.s. and moves all $Q^2$ blocks
to the left using the commutator $[Q^2,~]$, two types of terms will
be generated, (i) terms with $Q^2$ only inside commutators and (ii)
terms with one or more $Q^2$ blocks at the left. The terms of type (i)
are those corresponding to $\sum_n \tilde a_n\tau^n$. To lowest orders
one finds
\begin{eqnarray}
\tilde a_0 &=& 1 \,, \nonumber \\
\tilde a_1 &=& -M \,, \nonumber \\
\tilde a_2 &=& \frac{1}{2}M^2  -\frac{1}{6}M_{ii} 
+\frac{1}{12}F_{ij}^2
+\frac{1}{2}[Q^2,M]+\frac{1}{6}(Q^2)_{ii} \,.
\end{eqnarray}

Once the $\tilde a_n$ coefficients are so constructed one has to
proceed to rearrange (\ref{eq:18a}) as an expansion in powers of $M$,
$\D_i,$ and $\D_0$. The expansions in $M$ and $\D_i$ are already
inherited from (\ref{eq:18}). It remains to expand $[Q^2,~]$ in terms
of $[Q,~]$, or equivalently, in terms of $\D_0=[D_0,~]$ since the
quantities $Q$ and $D_0$ differ by a c-number. To do this, in the
$\tilde a_n$ coefficients $Q$ is to be moved to left, introducing
$\D_0$, until all the terms so generated are local operators made out
of $\D_\mu$ and $M$, and all uncommutated $Q$'s are at the left. E.g.
\begin{eqnarray}
\tilde a_2 &=& \frac{1}{2}M^2  -\frac{1}{6}M_{ii} 
+\frac{1}{12}F_{ij}^2
- \frac{1}{2}M_{00}
+ \frac{1}{3}E^2_i
+ \frac{1}{6}E_{0ii}
+ QM_0 
- \frac{1}{3}QE_{ii}
 \,.
\end{eqnarray}
(Recall that $E_i$ stands for the electric field $F_{0i}$.)  We can
see two types of contributions in $\tilde a_2$, namely, those without
a $Q$ at the left and those with one. If $Q$ is assigned an
engineering dimension of mass, all the terms are of the same
dimension, mass to the fourth. However, in our counting only the
dimension carried by $\D_\mu$ and $M$ is computed, and so the two
types of terms are of different order, namely, mass to the fourth and
mass to the third, respectively. Indeed, when $\tilde a_2$ is
introduced in (\ref{eq:18a}) (i.e. it gets multiplied by $e^{\tau
Q^2}$) and then in (\ref{eq:17}) (the sum over the Matsubara
frequencies is carried out) we will obtain the contributions (using
$\sum_{p_0}Q^n e^{\tau Q^2}\sim \varphi_n$)
\begin{equation}
\tilde a_2 \to \varphi_0(\Omega) \left( \frac{1}{2}M^2  -\frac{1}{6}M_{ii} 
+\frac{1}{12}F_{ij}^2
- \frac{1}{2}M_{00}
+ \frac{1}{3}E^2_i
+ \frac{1}{6}E_{0ii}
\right) \tau^2
+ \varphi_1(\Omega) \left(
 M_0 
- \frac{1}{3}E_{ii}
\right) \tau^{3/2} \,.
\end{equation}
These are contributions to the thermal heat kernel coefficients
$a^T_2$ and $a^T_{3/2}$, respectively, introduced in
(\ref{eq:18b}). Note the presence of half-integer order coefficients
from terms with an odd number of $Q$'s.

As we have just shown, each zero temperature heat kernel coefficient
$a_k$ in (\ref{eq:18}) allows to obtain a corresponding coefficient
$\tilde a_k$ with the same engineering dimension $2k$. Such
coefficient in turn contributes, in general, to several heat thermal
coefficients $a^T_n$ (with mass dimension $2n$). Let us discuss in
detail to which $a^T_n$ contributes each $\tilde a_k$.  The change
from engineering to real dimension comes about because some terms in
$\tilde a_k$ contain factors of $Q$ at the left which do no act as
$\D_0$ and so count as dimensionless. Therefore it is clear that for
given $k$, the allowed $n$ satisfy $n\le k$, the equal sign
corresponding to terms having all $Q$'s in commutators. On the other
hand, the maximum number of $[Q^2,~]$'s in $\tilde a_k$ ($k>0$) is
$k-1$, and from these, at most $k-1$ uncommutated $Q$'s can reach the
left of the term. This yields the further condition $k\le 2n-1$. Note
further that a factor $Q^\ell$ gives rise to a coefficient
$\varphi_\ell(\Omega)$ in $a^T_n$. In summary, in the computation of
the thermal coefficients $a^T_n$ up to $n=3$ (mass dimension 6), we
find the following scheme
\begin{eqnarray}
a_0 \sim \tilde a_0 &\sim &   \varphi_0 \, a^T_0 
\nonumber \\
a_1 \sim \tilde a_1 &\sim &   \varphi_0 \, a^T_1 
\nonumber \\
a_2 \sim \tilde a_2 &\sim &   \varphi_0 \, a^T_2 
                   + \varphi_1 \, a^T_{3/2}
\nonumber \\
a_3 \sim \tilde a_3 &\sim &   \varphi_0 \, a^T_3 
                   + \varphi_1 \, a^T_{5/2}
                   + \varphi_2 \, a^T_2
\nonumber \\
a_4 \sim \tilde a_4 &\sim &   \varphi_0 \, a^T_4 
                   + \varphi_1 \, a^T_{7/2}
                   + \varphi_2 \, a^T_3
                   + \varphi_3 \, a^T_{5/2}
\nonumber \\
a_5 \sim \tilde a_5 &\sim &   \varphi_0 \, a^T_5 
                   + \varphi_1 \, a^T_{9/2}
                   + \varphi_2 \, a^T_4
                   + \varphi_3 \, a^T_{7/2}
                   + \varphi_4 \, a^T_3
\label{eq:21}
\end{eqnarray}

The mixing of terms is a nuisance that does not occur at zero
temperature, however, it cannot be avoided: $Q$ contains $p_0$ and
must count as zeroth order (otherwise, if $Q$ were of order one the
expansion would consist of polynomials in $Q$ and the sum over $p_0$
would not converge). On the other hand, counting $p_0$ as zeroth order
and $D_0$ as first order even when it is inside $Q$ results in a
breaking of gauge invariance, as we noted at the end of the previous
subsection. The fact that $\Omega$ counts as dimensionless and $\D_0$
as dimension 1 is necessary to have an order by order gauge invariant
expansion. This counting is well defined provided that all $\Omega$'s
are at the left (for instance) of the local operators
(cf. (\ref{eq:33}) and discussion below).

From (\ref{eq:21}) we can see that we do not need the complete zero
temperature coefficients $a_4$ and $a_5$. $a^T_3$ requires only terms
$Y^n$, with $n=2,3,4$ in $a_4(Y,\D_i)$ and $n=4,5$ in $a_5(Y,\D_i)$.
We have extracted the zero temperature coefficients from
\cite{Bel'kov:1996tn}. These authors actually provide the traced
coefficients $b_n(x)$ defined by
\begin{equation}
\Tr\left(e^{-\tau(Y-D_i^2)}\right) =
(4\pi\tau)^{-d/2}\sum_{n=0}^\infty \int d^dx \,\tr\left(
b_n\right) \tau^n \,,
\label{eq:29}
\end{equation}
where $\Tr$ is the trace in the full Hilbert space of wave functions
and $\tr$ is the trace over the internal space only. The coefficient
$a_n$ is obtained by means of a first order variation of $b_{n+1}$
(cf. (\ref{eq:38})). The advantage of this procedure is that the
traced coefficients are much more compact and better checked.

As we have said, we have computed the thermal heat kernel coefficients
up to and including mass dimension 6 by the procedure just described
and also by that detailed in \cite{Garcia-Recio:2000gt}. This latter
approach uses the symbols method for space and time coordinates and so
computes the coefficients from scratch (in passing it yields the zero
temperature coefficients as well). We have verified that the two
computations give identical results after using the appropriate
Bianchi identities (in practice the method of
\cite{Garcia-Recio:2000gt} tends to give somewhat more compact
expressions). The results are as follows
\begin{eqnarray}
\ha_0 &=& \varphi_0 \,, 
\nonumber \\ 
\ha_{1/2} &=& 0 \,, 
\nonumber \\ 
\ha_1 &=& -\varphi_0 M  \,, 
\nonumber \\
\ha_{3/2} &=& \varphi_1 \left(M_0-\frac{1}{3}E_{ii}\right)  \,, 
\nonumber  \\
\ha_2 &=& 
\varphi_0 \, a_2^{T=0}
+ \frac{1}{6} \overline\varphi_2 (E_i^2+E_{0ii}-2M_{00}) 
 \,, 
\nonumber  \\
\ha_{5/2} &=& \frac{1}{3}\left(2\varphi_1+\varphi_3\right)M_{000}
+\frac{1}{6}\varphi_1 M_{0ii}-\frac{1}{3}\varphi_1 \left(2M_0 M + M M_0\right)
 \\
&& +\frac{1}{6}\varphi_1 \left ( \{M_i,E_i\}+ \{M,E_{ii}\}  \right)
-\left(\frac{1}{3}\varphi_1 + \frac{1}{5}\varphi_3 \right) E_{00ii}
-\frac{1}{30}\varphi_1 E_{iijj} 
\nonumber \\
&& -\left(\frac{5}{6}\varphi_1+\frac{2}{5}\varphi_3\right) E_{0i}E_i
-\left(\frac{1}{2}\varphi_1+\frac{4}{15}\varphi_3\right)E_i E_{0i}
+\frac{1}{30}\varphi_1 [ E_j , F_{iij}] 
\nonumber \\
&& -\varphi_1 \left(\frac{1}{10}F_{0ij}F_{ij}+\frac{1}{15}F_{ij} F_{0ij}\right)
 \,, 
\nonumber  \\
\ha_3 &=& \varphi_0 \, a_3^{T=0}
- \left(\frac{1}{4}\overline\varphi_2 - \frac{1}{10}\overline\varphi_4 \right)
M_{0000}
-\frac{1}{60}\overline\varphi_2 \Big( 
 3 M_{00ii}- 15 M_{00}M  - 5 M M_{00} - 15 M_0^2 
\nonumber \\
&&
+ 4 \{M,E_i^2\} + 2 E_i M E_i
+4 M E_{0ii}+6 E_{0ii}M
+ 4 M_i E_{0i}+ 6 E_{0i}M_i
\nonumber \\
&& 
+ 7 M_0 E_{ii}+ 3 E_{ii}M_0
+6 M_{0i}E_i+ 4 E_i M_{0i}
\Big)
\nonumber \\
&& 
+ \left(\frac{3}{20}\overline\varphi_2 - \frac{1}{15}\overline\varphi_4 \right)
 E_{000ii}
+\frac{1}{60}\overline\varphi_2 E_{0iijj}
+ \left(\frac{1}{2}\overline\varphi_2 - \frac{1}{5}\overline\varphi_4 \right)
E_{00i}E_i
\nonumber \\
&& 
+ \left(\frac{7}{30}\overline\varphi_2 - \frac{1}{10}\overline\varphi_4 \right)
E_i E_{00i}
+ \left(\frac{19}{30}\overline\varphi_2 - \frac{4}{15}\overline\varphi_4 \right)
E_{0i}^2
\nonumber \\
&& +\frac{1}{180}\overline\varphi_2
\Big(
2\{E_i,E_{jji}\}+4\{E_i,E_{ijj}\}+
5E_{ii}^2 + 4E_{ij}^2+4F_{0iij}E_j 
- 2E_j F_{0iij} - 2 E_{0ij}F_{ij}
\nonumber \\
&& 
-[E_{ij},F_{0ij}]-4E_{0i}F_{jji}
+2F_{jji}E_{0i}+2E_i F_{ij}E_j + 2\{E_i E_j,F_{ij}\}
+7F_{00ij}F_{ij}+3F_{ij}F_{00ij}+8F_{0ij}^2\Big)
 \,.
\nonumber
\end{eqnarray}
In these formulas $a_n^{T=0}$ stands for the zero temperature
coefficient. These are the same as those in (\ref{eq:20}) but using
$M$ instead of $Y$ and space-time indices instead of space indices,
e.g. $a_2^{T=0}= \frac{1}{2}M^2 -\frac{1}{6}M_{\mu\mu}
+\frac{1}{12}F_{\mu\nu}^2$. For convenience we have introduced the
auxiliary functions
\begin{equation}
\overline\varphi_2= \varphi_0+2\varphi_2 \,,\quad 
\overline\varphi_4= \varphi_0-\frac{4}{3}\varphi_4\,,
\quad
\overline\varphi_{2n}= \varphi_0-\frac{(-2)^n}{(2n-1)!!}\varphi_{2n}\,,
\label{eq:2.33}
\end{equation}
which vanish at $\tau/\beta^2=0$. Due to the Bianchi identity there is
some ambiguity in writing the terms. We have chosen to order the
derivatives so that all spatial derivatives are done first and the
temporal derivatives are the outer ones. This choice appears naturally
in our approach and in addition is optimal to obtain the traced
coefficients $b^T_n$ since the zeroth derivative of the Polyakov loop
vanishes (cf. (\ref{eq:33}) below), and so terms of the form
$\varphi_n X_0$ do not contribute to the traced coefficients upon
using integration by parts. The terms $\ha_0$, $\ha_1$, $\ha_{3/2}$
and $\ha_2$ were given in \cite{Megias:2002vr}.

\subsection{The traced thermal heat kernel coefficients}
\label{sec:2.c}

The zero temperature traced heat kernel coefficients have been
introduced in (\ref{eq:29}) (for the $d$-dimensional operator
$Y-D_i^2$). Of course, the choice $b_n=a_n$ would suffice, however,
exploiting the trace cyclic property and integration by parts more
compact choices are possible. At lowest orders the coefficients can be
taken as (we give the formulas for $K=M-D_\mu^2$ at zero temperature;
the heat kernel coefficients are dimension independent)
\cite{Bel'kov:1996tn,Fliegner:1998rk}
\begin{eqnarray}
b_0 &=& 1 \,, \nonumber \\
b_1 &=& -M \,, \nonumber \\
b_2 &=& \frac{1}{2}M^2+\frac{1}{12}F_{\mu\nu}^2
\,,  \label{eq:20a}  \\
b_3 &=& -\frac{1}{6}M^3
-\frac{1}{12}M_\mu^2
-\frac{1}{12}F_{\mu\nu}M F_{\mu\nu}
-\frac{1}{60}F_{\mu\mu\nu}^2
+\frac{1}{90}F_{\mu\nu}F_{\nu\alpha}F_{\alpha\mu}
 \,.
\nonumber
\end{eqnarray}
By construction $a_n-b_n$ is a commutator which vanishes inside $\Tr$.
Likewise, we can introduce the traced coefficients at finite
temperature,
\begin{equation}
\Tr\left(e^{-\tau(M-D_\mu^2)}\right) =
(4\pi\tau)^{-(d+1)/2}\sum_n \int d^{d+1}x\tr\left(
b^T_n\right) \tau^n \,,
\end{equation}
with $b^T_n$ simpler than $a^T_n$. Once again we choose a canonical
form for these coefficients where a function of $\Omega$ put at the
left is multiplied by a local operator (i.e. an operator made out of
$M$ and $\D_\mu$). To simplify the traced coefficients and bring them
to the canonical form we need to work out the commutators of the form
$[X,f(\Omega)]$ (in particular $\D_\mu f(\Omega)$) as a combination of
terms of the type function of $\Omega$ times local operator. As shown
in Appendix \ref{app:A}, the rules are as follows: let $f$ denote a
function of $\Omega$ (e.g.  $\varphi_n(\Omega)$) and let $f^{(n)}$ be
its $n$-th derivative with respect to the variable
$-\log(\Omega)/\beta$, then
\begin{eqnarray}
\D_0 f &=& 0 \,,
\nonumber \\
\D_i f &=& -f^\prime E_i
+\frac{1}{2}f^{\prime\prime} E_{0i}
-\frac{1}{3!}f^{(3)} E_{00i} 
+\cdots  \,, 
\label{eq:33} \\
{[X,f]} &=& -f^\prime X_0
+\frac{1}{2}f^{\prime\prime} X_{00}
-\frac{1}{3!}f^{(3)} X_{000}
+\cdots  \,.
\nonumber
\end{eqnarray}
These formulas imply that, unlike the zero temperature case, the
cyclic property mixes terms of different order at finite temperature.
This is because, as noted above, $\D_0$ has dimensions of mass whereas
$\Omega$ counts as dimensionless. So for instance, $\varphi_0(\Omega)$
is of order zero and $\D_i$ is of first order, yet $\D_i
\varphi_0(\Omega)$ contains terms of all orders, starting with
dimension 2. As we will discuss below, this implies that there is a
certain amount of freedom in the choice of the traced coefficients.
To apply these commutation rules to $a^T_n$ we further need the
relation
\begin{equation}
\varphi_n^\prime=\sqrt{\tau}(2\varphi_{n+1}+n\varphi_{n-1}) \,.
\end{equation}
Using these rules we can apply integration by parts and cyclic
property to the previously computed coefficients $a^T_n$ and choose a
more compact form for them valid inside the trace. In this way we
obtain, up to mass dimension 6,
\begin{eqnarray}
\hb_0 &=& \varphi_0 \,, \nonumber \\
\hb_{1/2} &=& 0  \,,\nonumber \\
\hb_1 &=& -\varphi_0 M  \,,
\nonumber \\ 
\hb_{3/2} &=& 0  \,, 
\label{eq:37}
 \\
\hb_2 &=& 
\varphi_0 b_2
-\frac{1}{6}\overline\varphi_2 E_i^2 
 \,,
\nonumber
 \\
\hb_{5/2} &=& 
-\frac{1}{6}\varphi_1 \{ M_i,E_i \} 
 \,,
\nonumber
 \\
\hb_3 &=&
\varphi_0 b_3
+\frac{1}{6}\overline\varphi_2 \left(
\frac{1}{2} M_0^2
+E_i M E_i
+\frac{1}{10} E_{ii}^2
+\frac{1}{10} F_{0ij}^2
-\frac{1}{5} E_i F_{ij} E_j
\right)
%\nonumber \\ &&
+\left(
\frac{1}{10}\overline\varphi_4
-\frac{1}{6}\overline\varphi_2
\right) E_{0i}^2
 \,.
\nonumber 
\end{eqnarray}
This is the main result of these Section, where the $\varphi_n$
functions are given in (\ref{eq:2.15}) and (\ref{eq:2.33}). In these
formulas the $b_n$ are the zero temperature coefficients given in
(\ref{eq:20a}). We note that the coefficient $\hb_3$ above is not
identical to that given in \cite{Megias:2002vr}. (The coefficient in
\cite{Megias:2002vr} corresponds to replace $\varphi_0 b_3$ above by
$\varphi_0 b_3^\prime$, where $b_3^\prime$ differs from $b_3$ in
(\ref{eq:20a}) by a cyclic permutation.) The two versions of $\hb_3$
differ by higher order terms. In what follows we use the coefficient
in (\ref{eq:37}).

Several remarks should be made about these expressions. Either at zero
or finite temperature there is an ambiguity in the choice of the
traced coefficients $b^T_n$, however, the ambiguity is essentially
larger at finite temperature. Indeed, writing the expansion as
\begin{equation}
\Tr\left(e^{-\tau(M-D_\mu^2)}\right) =
(4\pi\tau)^{-(d+1)/2}\sum_n B^T_n\tau^n\,,
\label{eq:36}
\end{equation}
we find that, although $b_n$ is ambiguous, $B^{T=0}_n$ is not. This is
because at zero temperature the expansion is tied to a series
expansion in powers of a parameter (say, $\tau$). At finite
temperature the expansion is not tied to a parameter (it is rather a
commutator expansion) and so the ambiguity exits not only for $b^T_n$
but also for $B^T_n$. For instance, $b^T_2$ above has been expressed
in terms of the coefficient $b_2$ given in (\ref{eq:20a}). Nothing
changes at zero temperature if we add $M_{\mu\mu}$ to $b_2$ since the
addition is a pure commutator, however, in $b_2^T$ it would mean to
add $\varphi_0 M_{\mu\mu}$ which is no longer a pure commutator,
thereby changing the functional $B^T_2$. In fact, $\varphi_0
M_{\mu\mu}$, which is formally of dimension 4, can be expressed as a
sum of terms of dimension 5 and higher, using integration by parts and
the commutation rules (\ref{eq:33}). So the concrete choice of $b^T_2$
affects the form of the higher orders, $b^T_{5/2}$, $b^T_3$, etc.

Taking into account this ambiguity, our criterion for choosing the
traced coefficients has been to recursively bring the $b^T_n$ to a
compact form. We observe that inside the trace (upon the applying the
commutation rules) $a^T_{3/2}$ is a sum of terms of dimension $4$ and
higher, so we choose $b^T_{3/2}=0$. Then $a^T_2$, augmented with the
terms generated from $a^T_{3/2}$, is brought to the most compact
form. This in turn produces higher order terms which are added to
$a^T_{5/2}$, and so on. Of course, this is not the only possibility,
since taken a $b^T_n$ to be simplest may imply a greater complication
in the higher order coefficients. For instance, as can be shown, it is
possible to arrange the expansion so that all half-order traced
coefficients vanish. E.g. $b^T_{5/2}$ can be removed at the cost of
complicating $b^T_2$.

It should be clear that the ambiguity in the expansion $B^T_n$ in
(\ref{eq:36}) does not affect its sum but only amounts to a
reorganization of the series. On the other hand the untraced
coefficients $a^T_n$ are not ambiguous: once brought to their
canonical form they are unique functionals of $M$ and $A_\mu$.

The heat kernel is symmetric under transposition of operators, the
$b^T_n$ have been chosen so that this mirror symmetry holds at each
order.

As is well-known \cite{Ball:1989xg} not only the $a^T_n$ allow to
obtain the $b^T_n$ but also the converse is true. By their definition
\begin{equation}
\langle x| e^{-\tau(M-D_\mu^2)} |x\rangle
=
-\frac{1}{\tau}\frac{\delta }
{\delta M(x)} \Tr\left(e^{-\tau(M-D_\mu^2)}\right)\,.
\end{equation}
Using the expansions in both sides, one finds at zero temperature
(using (\ref{eq:36}))
\begin{equation}
a^{T=0}_n(x)= 
-\frac{\delta B^{T=0}_{n+1} }{\delta M(x)}  \,.
\label{eq:38}
\end{equation}
At finite temperature, the variation of $b^T_k$ contributes not only
to $a^T_{k-1}$ but also to all higher order coefficients, in
general. So we have instead
\begin{equation}
a^T_n(x) \simeq
-\frac{\delta}{\delta M(x)}  \sum_{1\le k \le n+1} B^T_k \tau^{k-n-1}\,,
\end{equation}
where in the r.h.s. only the terms of dimension $2n$ are to be
retained and $k$ takes integer as well as half-integer values.  We
have checked our results by verifying that this relation holds for our
coefficients.

\section{The one-loop effective action of chiral QCD at high temperature}
\label{sec:3}

Here we will apply the thermal heat kernel expansion just derived to
obtain the one-loop effective action of QCD with massless quarks in
the high temperature region. We remark that the effective action we
are referring to is the standard one in quantum field theory, namely,
the classical generator of the one-particle irreducible diagrams. As a
consequence our classical fields may be time-dependent. The quantum
effective action in the sense of dimensional reduction
\cite{Appelquist:1981vg}, as an effective field theory for the static
modes is of great relevance in high temperature QCD and is also
discussed below, in Section \ref{sec:4}. We will use the background
field method, which preserves gauge invariance
\cite{Dewitt:1967ub}. The Euclidean action is
\begin{equation}
S=-\frac{1}{2\g^2}\int d^4x\,\tr(F_{\mu\nu}^2) 
+ \int d^4x \,\overline{q}\thruu{D} q \,.
\label{eq:42}
\end{equation}
$D_\mu= \partial_\mu+A_\mu$, with $A_\mu$ and
$F_{\mu\nu}=[D_\mu,D_\nu]$ antihermitian matrices of dimension
$N$. They belong to the fundamental representation of the Lie algebra
of the gauge group SU($N$).\footnote{Our point of view will be that
$A_\mu$ itself is the quantum field, independently of any particular
choice of basis in su($N$). So the coupling constant $\g$ is also
independent of that choice. Due to gauge invariance $A_\mu$ is not
renormalized.}

\subsection{Quark sector}
\label{subsec:3.a}

In this subsection we work out the quark contribution which is
somewhat simpler than the gluon contribution. (The latter requires the
use of the adjoint representation, introduction of ghost fields and
treatment of the infrared divergences.) Upon functional integration of
the quark fields, the partition function of the system picks up the
following factor from the quark sector
\begin{equation}
Z_q [A]= \Det(\thruu{D}\,)^{N_f}= \Det(\thruu{D}{}^2)^{N_f/2} \,,
\end{equation}
where $N_f$ denotes the number of quark
flavors. (As usual, we have squared the Dirac operator to obtain a
Klein-Gordon operator.) The corresponding contribution to the
effective action is (we use the convention $Z=e^{-\Gamma[A]}$)
\begin{equation}
\Gamma_q[A]= -\frac{N_f}{2}\Tr\,\log(\thruu{D}{}^2) =
\frac{N_f}{2}\int_0^\infty\frac{d\tau}{\tau}
\Tr \exp\left( \tau\thruu{D}{}^2\right)
=: \int d^4x \cL_q(x) \,,
\label{eq:3.3}
\end{equation}
\begin{equation}
\cL_q(x)= \frac{N_f}{2}\int_0^\infty\frac{d\tau}{\tau}
\frac{\mu^{2\epsilon}}{(4\pi\tau)^{D/2}}\sum_n\tau^n\tr(\hb_{n,q})
\,.
\label{eq:45}
\end{equation}
In this formula the Dirac trace is included in the $\hb_{n,q}$ and
$\tr$ refers to color trace (in the fundamental representation). The
ultraviolet divergences at $\tau=0$ are regulated using dimensional
regularization, with the convention $D= 4-2\epsilon$. As is standard
in dimensional regularization, the factor $\mu^{2\epsilon}$ is
introduced in order to deal with an effective Lagrangian of mass
dimension 4 rather than $4-2\epsilon$.

To apply our thermal heat kernel expansion we need only to identify
the corresponding Klein-Gordon operator. We use
\begin{equation}
\gamma_\mu=\gamma_\mu^\dagger\,,
\quad\gamma_\mu\gamma_\nu=\delta_{\mu\nu}+\sigma_{\mu\nu} \,, 
\quad \tr_{\text{Dirac}}(1)=4 \,.
\end{equation}
The expression
\begin{equation}
-\thruu{D}^2= -D_\mu^2-\frac{1}{2}\sigma_{\mu\nu}F_{\mu\nu}
\end{equation}
identifies $-\frac{1}{2}\sigma_{\mu\nu}F_{\mu\nu}$ as the (square)
mass term $M$ of the Klein-Gordon operator in this case. A direct
application of (\ref{eq:37}), shows that $\hb_1$ and $\hb_{5/2}$
cannot contribute (they have a single $M$ and this cancels due to the
trace over Dirac space). The other coefficients give, to mass
dimension 6 included,
\begin{eqnarray}
\hb_{0,q}&=& 4\varphi_0 \,, \nonumber \\ 
\hb_{2,q} &=& -\frac{2}{3}\left(\varphi_0 F_{\mu\nu}^2
+ \overline\varphi_2
E_i^2 \right) \,,
 \\ 
\hb_{3,q} &=&
\varphi_0 \left(
\frac{32}{45}F_{\mu\nu}F_{\nu\lambda}F_{\lambda\mu}
+\frac{1}{6}F_{\lambda\mu\nu}^2 
-\frac{1}{15}F_{\mu\mu\nu}^2
\right)
+\overline\varphi_2\left(
\frac{1}{15}E_{ii}^2-\frac{1}{10}F_{0ij}^2-\frac{2}{15}E_i F_{ij}E_j
\right)
%\nonumber \\ && 
+\left(\frac{2}{5}\overline\varphi_4-\overline\varphi_2 \right) E_{0i}^2
\,. \nonumber
\end{eqnarray}
In these formulas the functions $\varphi_n$ (defined in
(\ref{eq:2.15}) and (\ref{eq:2.33})) correspond to their fermionic
versions. All fields are in the fundamental representation.

The required integrals over $\tau$ in (\ref{eq:45}) are of the form
\begin{equation}
I^\pm_{\ell,n}(\omega) :=
\int_0^\infty\frac{d\tau}{\tau}(4\pi\mu^2\tau)^\epsilon 
\tau^\ell\varphi^\pm_n(\omega)\,, \quad |\omega|=1 
\label{eq:3.8}
\end{equation}
where $\varphi^\pm_n$ refers to the bosonic or fermionic version,
respectively. In the quark sector the argument $\omega$ will be the
Polyakov loop in the fundamental representation, or, in practice, any
of its eigenvalues. These integrals can be done in closed form (see
Appendix \ref{app:B}). In particular
\begin{eqnarray}
I^-_{\ell,2n}(e^{2\pi i\nu})
&=& (-1)^n (4\pi)^\epsilon \left(\frac{\mu\beta}{2\pi}\right)^{2\epsilon}
\left(\frac{\beta}{2\pi}\right)^{2\ell} 
\frac{\Gamma(\ell+n+\epsilon+\frac{1}{2})}
{\Gamma(\frac{1}{2})}
\Big[ \zeta(1+2\ell+2\epsilon,\half+\nu)
+\zeta(1+2\ell+2\epsilon,\half-\nu) \Big] \,, \nonumber \\ 
 && \quad
-\frac{1}{2} < \nu < \frac{1}{2} \,.
\label{eq:50}
\end{eqnarray}
The integrals $I^\pm_{\ell,n}(\omega)$ are one-valued functions of
$\omega$, i.e. periodic in terms of $\nu$, however, to apply the
explicit formula (\ref{eq:50}) $\nu$ has to be taken in the interval
$-\frac{1}{2} < \nu < \frac{1}{2}$. The generalized Riemann
$\zeta$-function $\zeta(z,q)=\sum_{n=0}^\infty(n+q)^{-z}$ has only a
single pole at $z=1$ \cite{Gradshteyn:1980bk}, so the dimensionally
regulated integrals yield the standard pole of the type $1/\epsilon$
solely for the integrals $I^-_{0,2n}$, which appear in $\hb_{2,q}$.

We can now proceed to compute the contributions to the effective
Lagrangian. The zeroth order requires $I^-_{-2,0}$. Using the relation
$\zeta(1-n,q)= -B_n(q)/n$, $n= 1,2,\ldots$, with $B_n(q)$ the
Bernoulli polynomial of order $n$ \cite{Gradshteyn:1980bk}, one finds
\begin{equation}
I^-_{-2,0}=
-\frac{2}{3}\left(\frac{2\pi}{\beta}\right)^4B_4(\half+\nu)
+O(\epsilon)
\end{equation}
so the effective potential is
\begin{equation}
\cL_{0,q}(x)= \pi^2N_f T^4 \left(
\frac{2N}{45}-\frac{1}{12}\tr\left[(1-4\overline{\nu}^2)^2\right] \right)
\,,\quad \Omega(x)=e^{2\pi i\overline{\nu}}\,,\quad -\frac{1}{2} < \overline{\nu} < \frac{1}{2} \,.
\label{eq:3.11}
\end{equation}
Here $N$ is the number of colors, $\tr$ is taken in the fundamental
representation of the gauge group and $\overline{\nu}$ is the matrix
$\log(\Omega)/(2\pi i)$ with eigenvalues in the branch
$|\overline{\nu}|<1/2$. This is the well known result \cite{Gross:1981br}.

The terms of mass dimension 4 have a pole at $\epsilon=0$. Using the
relation
\begin{equation}
\zeta(1+z,q)= \frac{1}{z}-\psi(q)+O(z)\,,
\end{equation}
(where $\psi(q)$ is the digamma function) one finds
\begin{eqnarray}
I^-_{0,0} &=&
\frac{1}{\epsilon}+\log(4\pi)-\gamma_E + 2\log(\mu\beta/4\pi)
-\psi(\half+\nu)-\psi(\half-\nu)+O(\epsilon)\,, 
\nonumber \\
I^-_{0,\overline{2}} &:=& I^-_{0,0}+2I^-_{0,2}= -2 +O(\epsilon) \,.
\label{eq:3.13}
\end{eqnarray}
(For convenience, we have introduced the integrals
$I^{\pm}_{\ell,\overline{2n}}$ analogous to $I^{\pm}_{\ell,2n}$ in
(\ref{eq:3.8}) but using $\overline\varphi_{2n}$ instead of
$\varphi_{2n}$.)

The terms $(\epsilon^{-1}+\log(4\pi)-\gamma_E)$ in $I^-_{0,0}$ come
with $\tr(F_{\mu\nu}^2)$ and are removed by adopting the
$\overline{\text{MS}}$ scheme. We will discuss this in conjunction
with gluon sector. After renormalization
\begin{equation}
\cL_{2,q}(x)= -\frac{1}{3}\frac{1}{(4\pi)^2} N_f \tr \left[
\left(2\log(\mu/4\pi T)
-\psi(\half+\overline{\nu})-\psi(\half-\overline{\nu})
\right)F_{\mu\nu}^2 -2E_i^2\right] \,.
\label{eq:14}
\end{equation}

Finally, the terms of mass dimension 6 in four space-time dimensions
require $I^-_{1,0}$, $I^-_{1,\overline{2}}$, and
$I^-_{1,\overline{4}}$. Using the relation
$\psi^{(n)}(q)=(-1)^{n+1}n!\zeta(n+1,q)$ \cite{Gradshteyn:1980bk}, one
obtains
\begin{equation}
I^-_{1,0}= -\left(\frac{\beta}{4\pi}\right)^2 \left(
\psi^{\prime\prime}(\half+\nu) +\psi^{\prime\prime}(\half-\nu) \right)
 +O(\epsilon) \,, \quad
I^-_{1,\overline{2}}= -2 I^-_{1,0} +O(\epsilon)
 \,, \quad
I^-_{1,\overline{4}}= -4 I^-_{1,0} +O(\epsilon)
\,.
\label{eq:56}
\end{equation}
(All these integrals are related through simple proportionality
factors, as follows from (\ref{eq:B6})). This yields
\begin{eqnarray}
\cL_{3,q}(x) &=& -\frac{2}{(4\pi)^4 } \frac{N_f}{T^2}
 \tr \bigg[
\Big(\psi^{\prime\prime}(\half+\overline{\nu}) +\psi^{\prime\prime}(\half-\overline{\nu}) \Big)
\nonumber  \\
&& \times
\left( 
\frac{8}{45}F_{\mu\nu}F_{\nu\lambda}F_{\lambda\mu}
+\frac{1}{24}F_{\lambda\mu\nu}^2
-\frac{1}{60}F_{\mu\mu\nu}^2
+\frac{1}{20}F_{0\mu\nu}^2
%+\frac{1}{10}E_{0i}^2
-\frac{1}{30}E_{ii}^2
%+\frac{1}{20}F_{0ij}^2
+\frac{1}{15}E_iF_{ij}E_j
\right)
\bigg]
\,.
\label{eq:16}
\end{eqnarray}
In all these formulas $\overline{\nu}$ is the matrix
$\log(\Omega)/(2\pi i)$ in the fundamental representation and in the
branch $|\overline{\nu}|<\frac{1}{2}$ in the eigenvalue sense.  Note
the hierarchy in powers of temperature, $\cL_0\sim T^4$, $\cL_2\sim
T^0$, $\cL_3\sim T^{-2}$, implying that the heat kernel expansion at
finite temperature is essentially an expansion on $k^2/T^2$ with $k$
the typical gluon momentum. Terms of order $T^2$ are forbidden since
there is no available gauge invariant operator of dimension two.

\subsection{Gluon sector}
\label{subsec:3.b}

In the background field approach \cite{Dewitt:1967ub} the gluon field
is split into a classical field plus a quantum fluctuation,
i.e. $A_\mu \to A_\mu+a_\mu$ in the action (\ref{eq:42}). As is
standard in the effective action formalism, the appropriate currents
are added so that the classical field $A_\mu$ is a solution of the
equations of motion (and so no terms linear in the fluctuation
remain). The one-loop effective action corresponds then to neglect
contributions beyond the quadratic terms in the quantum fluctuations
and integrate over $a_\mu$. (The quark fields are taken as pure
fluctuation, so $a_\mu$ does not change the quark sector at one-loop.)

The quadratic piece of the gluon action is
\begin{equation}
S^{(2)}= -\frac{1}{\g^2} \int d^4x\, \tr\left[
-a_\nu\D_\mu^2 a_\nu -2 \, a_\mu[F_{\mu\nu},a_\nu]-(\D_\mu a_\mu)^2 
\right] \,.
\label{eq:58}
\end{equation}
Here all covariant derivatives are those associated to the classical
gluon field $A_\mu$. Note that the first two terms are of the
standard Klein-Gordon form, but the last one is not. Before doing the
functional integration over $a_\mu$ one has to fix the gauge of these
fields. This implies adding a gauge fixing term and the corresponding
Faddeev-Popov term \cite{Bornsen:2002hh} in the action. We take the
covariant Feynman gauge $\D_\mu a_\mu=f(x)$, since the associated
gauge fixing action precisely cancels the offending term $(\D_\mu
a_\mu)^2$ in (\ref{eq:58}). After adding the ghost term one has
\begin{equation}
S^{(2)}= -\frac{1}{\g^2} \int d^4x\, \tr\left[
-a_\nu\D_\mu^2 a_\nu -2 \, a_\mu[F_{\mu\nu},a_\nu] - \bar{C}\D_\mu^2 C
\right] \,.
\label{eq:58a}
\end{equation}
The coupling constant has no effect here since it can be absorbed in
the normalization of the fields. The ghost fields $C$ and $\bar{C}$
are anticommuting (although periodic in Euclidean time) and are
matrices in the fundamental representation of su($N$).

The full effective action (to one loop) is
\begin{equation}
\Gamma[A]= -\frac{\mu^{-2\epsilon}}{2\g^2_0}\int
d^Dx\,\tr(F_{\mu\nu}^2) + \Gamma_q[A]+\Gamma_g[A] \,,
\end{equation}
where the first piece is the tree level action (accounting for
renormalization; $\g_0$ is dimensionless), the second one is the quark
contribution, obtained in the previous subsection, and the last term
follows from functional integration over $a_\mu$ and $C$, $\bar{C}$ in
(\ref{eq:58a}),
\begin{equation}
\Gamma_g[A]= \frac{1}{2}\Tr\log\left(-\D_\mu^2-2\F_{\mu\nu} \right)
-\Tr\log\left(-\D_\mu^2\right)
=: \int d^4x \cL_g(x)\,.
\end{equation}
where $\D_\mu= [D_\mu,~]$ and $\F_{\mu\nu}= [F_{\mu\nu},~]$. From
(\ref{eq:58a}), we can see that the Klein-Gordon operator over the
gluon field $a_\mu$ acts on an internal space of dimension $D\times
(N^2-1)$, where $D=4-2\epsilon$ is the number of gluon polarizations
(including the two unphysical ones) and corresponds to the Lorentz
index $\mu$, and $N^2-1$ is the dimension of the adjoint
representation of the group. $\D_\mu$ and $\F_{\mu\nu}$ act in the
adjoint representation. The covariant derivative of the Klein-Gordon
operator is the identity in the Lorentz space whereas the ``mass
term'' is a matrix in that space, namely, $(M)_{\mu\nu}=
-2\F_{\mu\nu}$. Similarly, the space of the Klein-Gordon operator over
the ghost fields has dimension $N^2-1$, the mass term is zero and
the corresponding covariant derivative is just $D_\mu$ but in the
adjoint representation.

Applying once again the heat kernel representation, we have
\begin{equation}
\cL_g(x)= -\frac{1}{2}\int_0^\infty\frac{d\tau}{\tau}
\frac{\mu^{2\epsilon}}{(4\pi\tau)^{D/2}}\sum_n\tau^n
\htr(\hb_{n,g}) \,,
\label{eq:45g}
\end{equation}
where for convenience, the Lorentz trace over gluons as well as the
ghost contribution are included in the coefficient
$\hb_{n,g}$. $\htr$ denotes the color trace in the adjoint
representation. A straightforward calculation yields
\begin{eqnarray}
\hb_{0,g}&=& (D-2)\varphi_0(\Om)  \,, \nonumber \\ 
\hb_{2,g} &=& \left(-2+\frac{D-2}{12}\right)
\varphi_0(\Om) \F_{\mu\nu}^2
-\frac{D-2}{6} \overline\varphi_2(\Om)
\E_i^2 \,,
 \\ 
\hb_{3,g} &=& 
\varphi_0(\Om)  \left(
\left(\frac{4}{3}+\frac{D-2}{90} \right) 
\F_{\mu\nu}\F_{\nu\lambda}\F_{\lambda\mu}
+\frac{1}{3}\F_{\lambda\mu\nu}^2 
-\frac{D-2}{60}\F_{\mu\mu\nu}^2
\right) \nonumber \\
&& 
+ \, 
\frac{1}{6}\overline\varphi_2(\Om) \left(
-2\F_{0\mu\nu}^2 + 
\frac{D-2}{10}\left( \E_{ii}^2+\F_{0ij}^2-2 \E_i \F_{ij}\E_j \right)\right)
\nonumber \\
&& 
+ \, 
(D-2)\left(\frac{1}{10}\overline\varphi_4(\Om) - \frac{1}{6}
\overline\varphi_2(\Om) \right) \E_{0i}^2
\,.
 \nonumber 
\end{eqnarray}
The coefficients $\hb_{1,g}$ and $\hb_{5/2,g}$ vanish, as do all terms
with a single $M$, due to the Lorentz trace. The contributions with
$D-2$ come from pieces without $M$ in (\ref{eq:37}) and
(\ref{eq:20a}). The effect of the ghost is to remove two gluon
polarizations, $D\to D-2$. Unlike the fermionic case, the thermal heat
kernel coefficients depend explicitly on the space-time dimension
through these polarization factors. In these formulas the functions
$\varphi_n$ correspond to their bosonic versions. In addition, its
argument $\Om$ and all field strength tensor and covariant derivatives
are in the adjoint representation.

We can now proceed to the calculation of the effective Lagrangian. We
note that the integrals over $\tau$ are no different to those for the
quark sector (see (\ref{eq:3.8}) and Appendix \ref{app:D}), after the
replacement $\nu \to \nu-\frac{1}{2}$ (coming from
$\varphi_n^+(\omega)= \varphi_n^-(-\omega)$) and so $0<\nu<1$ now:
\begin{eqnarray}
I^+_{\ell,2n}(e^{2\pi i\nu})
&=& (-1)^n (4\pi)^\epsilon \left(\frac{\mu\beta}{2\pi}\right)^{2\epsilon}
\left(\frac{\beta}{2\pi}\right)^{2\ell} 
\frac{\Gamma(\ell+n+\epsilon+\frac{1}{2})}
{\Gamma(\frac{1}{2})}
\Big[ \zeta(1+2\ell+2\epsilon,\nu)
+\zeta(1+2\ell+2\epsilon,1-\nu) \Big] \,, \nonumber \\ 
 && \quad
0 < \nu < 1 \,.
\label{eq:50b}
\end{eqnarray}

In this way, for the effective potential one obtains
\begin{eqnarray}
\cL_{0,g}(x) &=& \frac{\pi^2}{3} T^4 
\htr\left[B_4(\hnu)+B_4(1-\hnu)\right]
\label{eq:65} \\
&=&
-\frac{\pi^2}{45} T^4 \, (N^2-1)
+ \frac{2\pi^2}{3} T^4 \,\htr\left[ \hnu^2(1-\hnu)^2 \right]
\,,\quad \hnu= \log(\Om)/(2\pi i)\,,\quad
0<\hnu<1\,.
\end{eqnarray}
This is also in agreement with the well known result
\cite{Gross:1981br}. We emphasize that $\Om$ and $\hnu$ are now in the
adjoint representation as indicated by the notation $\htr$.

The mass dimension 4 piece of the effective Lagrangian, coming from
$\hb_{2,g}$, requires $I^+_{0,0}$ which is ultraviolet divergent and
$I^+_{0,\overline{2}}$ which is UV finite (cf. (\ref{eq:3.13})). The
finite pieces, in the $\overline{\text{MS}}$ scheme, are found to be
\begin{equation}
\cL_{2,g}(x) =
\frac{1}{(4\pi)^2} \htr \Big[
\frac{11}{12} \left( 2\log(\mu/4\pi T) +\frac{1}{11}
-\psi(\hnu)-\psi(1-\hnu) \right) \F_{\mu\nu}^2
-\frac{1}{3} \E_i^2 
\Big]\,,\quad 0 < \hnu < 1 \,.
\label{eq:69}
\end{equation}

On the other hand, the divergent contribution in the gluon sector,
combined with that in the quark sector and the tree level Lagrangian
yields (all terms have been multiplied by the factor $\mu^{2\epsilon}$
to restore dimensions)
\begin{eqnarray}
\cL_{\text{tree}}(x)+
\cL^{\text{div}}_q(x)+\cL^{\text{div}}_g(x)
&=&
-\frac{1}{2\g^2_0}\,\tr(F_{\mu\nu}^2) + 
\frac{1}{(4\pi)^2}\left(\frac{1}{\epsilon}+\log(4\pi)-\gamma_E \right)
\left(
\frac{11}{12}\htr(\F_{\mu\nu}^2) 
-\frac{N_f}{3}\tr(F_{\mu\nu}^2) 
\right).
\label{eq:3.27}
\end{eqnarray}
Use of the SU($N$) identity (\ref{eq:3.28}) yields the renormalized
tree level Lagrangian
\begin{equation}
\cL_{\text{tree}}(x)+
\cL^{\text{div}}_q(x)+\cL^{\text{div}}_g(x)
=
-\frac{1}{2\g^2(\mu)}\,\tr(F_{\mu\nu}^2) \,,
\label{eq:3.29}
\end{equation}
with the standard one-loop renormalization group improved in the
$\overline{\text{MS}}$ scheme
\begin{eqnarray}
\frac{1}{\g^2(\mu)}
=
\frac{1}{\g^2_0}- \beta_0
\left(\frac{1}{\epsilon}+\log(4\pi)-\gamma_E \right) \,,
\qquad \beta_0 = \frac{1}{(4\pi)^2}
\left( \frac{11}{3} N -\frac{2}{3} N_f \right)\,,
\end{eqnarray}
guarantying the scale independence of (\ref{eq:3.27}).
Note that, due to gauge invariance, the classical fields $A_\mu$ do
not need ultraviolet renormalization. (In the context of the
dimensionally reduced effective theory, finite, temperature dependent,
renormalization has been found to be useful in practice
\cite{Landsman:1989be,Chapman:1994vk}. See Section \ref{sec:4}.)

Putting together all terms of mass dimension 4 (renormalized tree
level plus one-loop), we find
\begin{eqnarray}
\cL_2(x) &=&
\left(-\frac{1}{2\g^2(\mu)} +\beta_0 \log(\mu/4\pi T)
+\frac{1}{6}\frac{1}{(4\pi)^2}N \right) \,\tr(F_{\mu\nu}^2) 
\nonumber
\\
&& -\frac{11}{12}\frac{1}{(4\pi)^2} \htr \left[
 \big( \psi(\hnu)+\psi(1-\hnu) \big)
\F_{\mu\nu}^2
\right]
\nonumber \\
&&
+\frac{1}{3}\frac{1}{(4\pi)^2} N_f \tr \left[
\left(\psi(\half+\bnu)+\psi(\half-\bnu) \right)F_{\mu\nu}^2
\right]
\nonumber \\
&&
-\frac{2}{3}(N-N_f)\frac{1}{(4\pi)^2}  
\tr \left[ E_i^2 \right]
\,,\qquad -\frac{1}{2} < \bnu < \frac{1}{2}
\,,\quad 0 < \hnu < 1 \,.
\label{eq:3.31}
\end{eqnarray}

The terms of mass dimension 6 are easily obtained from the coefficient
$\hb_{3,g}$ and the integrals $I^+_{1,0}$, $I^+_{1,\overline{2}}$ and
$I^+_{1,\overline{4}}$,
\begin{eqnarray}
\cL_{3,g}(x) &=&  \frac{1}{2}\frac{1}{(4\pi)^4 } \frac{1}{T^2}
\htr \bigg[
\Big(\psi^{\prime\prime}(\hnu) 
+ \psi^{\prime\prime}(1-\hnu) \Big)
\nonumber  \\
&& \times
\left( 
\frac{61}{45}\F_{\mu\nu}\F_{\nu\lambda}\F_{\lambda\mu}
+\frac{1}{3}\F_{\lambda\mu\nu}^2
-\frac{1}{30}\F_{\mu\mu\nu}^2
+\frac{3}{5}\F_{0\mu\nu}^2
%+\frac{6}{5}\E_{0i}^2
-\frac{1}{15}\E_{ii}^2
%+\frac{3}{5}\F_{0ij}^2
+\frac{2}{15}\E_i\F_{ij}\E_j
\right)
\bigg]
\,.
\end{eqnarray}
Note again the hierarchy in powers of temperature, $\cL_0\sim T^4$,
$\cL_2\sim T^0$, $\cL_3\sim T^{-2}$.

\subsection{Infrared divergence and other renormalization schemes}
\label{subsec:3.c}

The integrals $I^\pm_{\ell,n}$ may contain not only ultraviolet
divergences but also infrared ones (corresponding to the large $\tau$
region). Specifically, this happens if $\ell\ge 0$, $n=0$ and $e^{2\pi
i\nu}=\pm 1$ (see Appendix \ref{app:B}). In the quark sector (i.e., in
the fundamental representation), and for a generic configuration of
$A_0(x)$, no eigenvalue of $\Omega$ will be $-1$ in the bulk and so
such divergence can be disregarded. Unfortunately, in the gluon sector
the situation is different since for any gauge configuration at least
$N-1$ eigenvalues of $\Om(x)$ are necessarily unity. Therefore, the
singular value $\nu=$integer always appears when evaluating the
adjoint trace in $\cL_{2,g}$ and $\cL_{3,g}$. The infrared divergences
are characteristic of massless theories at finite temperature
\cite{Linde:1980ts,Braaten:1995na}.

For $\nu=0$, the infrared divergence comes solely from the static
Matsubara mode, $p_0=0$, in $\varphi_0$. The corresponding integral
over $\tau$ has no natural scale and so the point of view can be taken
that such divergences are automatically removed by dimensional
regularization \cite{Collins:1984bk}. As explained in Appendix
\ref{app:B}, the integrals $I^+_{\ell,2n}$ without the static mode are
given by the same expressions (\ref{eq:50b}) after the replacement
$\nu\to 1+\nu$ in the first $\zeta$-function. The resulting
prescription is then to use the formulas of $\cL_{2,g}$ and
$\cL_{3,g}$ with the replacements
\begin{eqnarray}
&& \psi(\hnu)+\psi(1-\hnu)\,\big|_{\hnu=0} \to
\psi(1+\hnu)+\psi(1-\hnu) \,\big|_{\hnu=0}
= -2\gamma_E \,,
\nonumber \\
&& \psi^{\prime\prime}(\hnu)+\psi^{\prime\prime}(1-\hnu)\,\big|_{\hnu=0} \to
\psi^{\prime\prime}(1+\hnu)+\psi^{\prime\prime}(1-\hnu) \,\big|_{\hnu=0}
= -4\zeta(3)\,,
\end{eqnarray}
to be made in the subspace $\Om=1$ only, when taking the trace in the
adjoint representation. One may worry that subtracting this subspace
is not consistent with gauge invariance. This is not so. As will be
discussed below, the periodicity of the effective action as a function
of $\log(\Om)$ is an important requirement. This property is not
spoiled by the previous prescriptions.

Alternatively, one can regulate the infrared divergence by including a
cutoff function $e^{-m^2\tau}$ in the $\tau$ integral. The infrared
finite modes are unaffected in the limit of small $m$. The static mode
in $\varphi_0$ develops power-like divergences to be added to the
result obtained through dimensional regularization. These terms are
easily computed and are\footnote{Note that $I^+_{\ell,\overline{2n}}$
also contains $\varphi_0$ and so is also afflicted by the
divergence. This implies that introducing $e^{-m^2\tau}$ is not
equivalent to a regularization of the digamma function (and its
derivatives) in the final formulas, since simple scaling relations of
the type (\ref{eq:56}) or (\ref{eq:B6}) no longer hold.}
\begin{eqnarray}
\cL_{2,\text{IR}} &=&
\frac{1}{48\pi}\frac{T}{m}\tr\left[ 11 F^2_{\mu\nu\perp}
+2E^2_{i\perp} 
\right] \,,
\nonumber  \\
\cL_{3,\text{IR}} &=&
\frac{1}{240\pi}\frac{T}{m^3}\tr\Big[ 
-\frac{61}{3} F_{\mu\nu\perp}F_{\nu\alpha}F_{\alpha\mu}
+E_{i\perp}F_{ij}E_{j\perp} 
%-E_{i\parallel}F_{ij}E_{j\parallel}  
+E_{i}F_{ij\parallel}E_{j}
\label{eq:3.33} \\ &&
-5 F^2_{\mu\nu\lambda\perp}
+\frac{1}{2}F^2_{\mu\mu\nu\perp}
+\frac{9}{2}F^2_{0\mu\nu\perp}
+3 E^2_{0i\perp}
-\frac{1}{2} E^2_{ii\perp}
\Big] \,.
\nonumber
\end{eqnarray}
Even though this is a gluonic term, the result has been expressed in
the fundamental representation, which is often
preferable. (Unfortunately this is not so easily done for the other
gluonic contributions, for a general SU($N$) group, due to the
presence of the Polyakov loop in the formulas.) In these expressions
we have used the notation $F_{\mu\nu\parallel}$ to denote the pieces
of $F_{\mu\nu}$ which commute with $\Omega$ and $F_{\mu\nu\perp}$ for
the remainder.  Specifically, in the gauge in which $\Omega$ is
diagonal, $F_{\mu\nu\parallel}$ is the diagonal part of
$F_{\mu\nu}$. As shown in Appendix \ref{app:D}, only terms involving
at least one perpendicular component may be infrared divergent, and
this is verified in (\ref{eq:3.33}).

We have used here the $\overline{\text{MS}}$ scheme in dimensional
regularization. Alternatively one can use Pauli-Villars regularization
which amounts to inserting a regulating factor $(1-e^{-\tau M^2})$ in
the $\tau$ integration \cite{Diakonov:2003yy}. All convergent
integrals (including $I^\pm_{0,\overline{2}}$) are unchanged in the
limit of large $M$, whereas
\begin{eqnarray}
I^{+,\text{PV}}_{0,0} &=&
 2\log(M/\mu)  +
 2\log(\mu\beta/4\pi)
-\psi(\nu)-\psi(1-\nu)+O(M^{-1})\,, \quad 0<\nu<1 \,,
\nonumber  \\
I^{-,\text{PV}}_{0,0} &=&
 2\log(M/\mu)  +
 2\log(\mu\beta/4\pi)
-\psi(\half+\nu)-\psi(\half-\nu)+O(M^{-1})\,, 
\quad -\frac{1}{2}<\nu<\frac{1}{2} \,.
\end{eqnarray}
(Note that these formulas do not actually depend on the scale $\mu$.)
The Pauli-Villars renormalized result is obtained by combining
$\log(M^2/\mu^2)$ with the bare coupling constant in the tree level
Lagrangian to yield the renormalized coupling constant
$\g_{\text{PV}}(\mu)$. If, as usual, the $\Lambda_R$ parameter in the
scheme $R$ is defined as the scale $\mu=\Lambda_R$ for which
$1/\g_R^2(\mu)$ vanishes, it is found that the Pauli-Villars and
$\overline{\text{MS}}$ schemes give identical renormalized results, at
one-loop, when
\begin{equation}
\log\left(\Lambda_{\text{PV}}^2/\Lambda_{\overline{\text{MS}}}^2
\right) = \frac{1}{11-2N_f/N} \,.
\end{equation}
The difference between both scales comes from the $\frac{1}{11}$ in
(\ref{eq:69}), which is due to the $-2\epsilon$ extra gluon
polarizations in the dimensional regularization scheme
\cite{Hasenfratz:1981tw}.

\subsection{Results for SU(2) and SU(3)}
\label{subsec:3.d}

We can particularize our formulas for SU(2) by working out the color
traces explicitly. We use the antihermitian su(2) basis
$\vec{\sigma}/2i$, so
\begin{equation}
A_0=-\frac{i}{2}\vec{\sigma}\cdot\vec{A}_0\,,\quad
F_{\mu\nu}=-\frac{i}{2}\vec{\sigma}\cdot\vec{F}_{\mu\nu}\,,
\quad\text{etc.}
\end{equation}
It is convenient to choose the ``Polyakov gauge'', in which $A_0$ is
time independent and diagonal \cite{Salcedo:1998sv}. In SU(2),
$A_0=-\frac{1}{2}i \sigma_3 \phi$. In this case the eigenvalues of the
Polyakov loop in the fundamental representation are $\exp(\pm i\beta
\phi/2)$, and in the adjoint representation are $\exp(\pm i\beta
\phi)$ and $1$. Full results for $\cL_{0,2,3}(x)$ in both sectors
are given in Appendix \ref{app:C}. Here we quote the results for
$\cL_2(x)$ from the gluon and quark loops:
\begin{eqnarray}
\cL_{2,q}(x) &=&
\frac{N_f}{48 \pi^2} \left[
\left( 2\log\left(\frac{\mu}{4\pi T}\right)
-\psi(\half+\bnu)-\psi(\half-\bnu)
-1 
\right) \vec{E}_i^2 
\right.
\nonumber \\ &&
 \left. +
\left(
2\log\left(\frac{\mu}{4\pi T}\right) 
-\psi(\half+\bnu)-\psi(\half-\bnu)
\right) \vec{B}_i^2
\right]
\,,
\end{eqnarray}
with $\bnu = (\beta\phi/4\pi+1/2) \;(\mod 1) -1/2$ and
$B_i=\half\epsilon_{ijk}F_{jk}$ is the magnetic field.

\begin{eqnarray}
\cL_{2,g}(x) &=&
-\frac{11}{48\pi^2}\left[
\left(
2\log(\mu/4\pi T)-\frac{1}{11}
-\psi(\hnu)-\psi(1-\hnu)
\right) \vec E^2_{i\parallel}
\right.
\nonumber \\
&& 
 +
\left(
\frac{12}{11}\frac{\pi T}{m}+2\log(\mu/4\pi T)-\frac{1}{11} +\gamma_E
-\half\psi(\hnu)-\half\psi(1-\hnu)
\right) \vec E^2_{i\perp}
\nonumber \\
&&
 +
\left(
2\log(\mu/4\pi T)+\frac{1}{11}
-\psi(\hnu)-\psi(1-\hnu)
\right) \vec B^2_{i\parallel}
\nonumber \\
&& 
\left.
 +
\left(
\frac{\pi T}{m}+2\log(\mu/4\pi T)+\frac{1}{11} +\gamma_E
-\half\psi(\hnu)-\half\psi(1-\hnu)
\right) \vec B^2_{i\perp}
\right] \,.
\end{eqnarray}
Here $\hnu= \beta\phi/2\pi ~(\mod 1)$, and
\begin{equation}
\vec E_i=\vec E_{i\parallel}+\vec E_{i\perp} \,, \quad
\vec B_i=\vec B_{i\parallel}+\vec B_{i\perp} \,,
\label{eq:3.40}
\end{equation}
are the decompositions of the electric and magnetic fields in the
directions parallel and perpendicular to $\vec A_0$. Such a
decomposition is gauge invariant.

The quark and gluon sector contributions are periodic in $\phi$ with
periods $4\pi T$ and $2\pi T$ respectively. This periodicity in $A_0$
of the coefficients multiplying the local operators is a consequence
of gauge invariance. Indeed, after choosing the Polyakov gauge there
is still freedom to make further non stationary gauge transformations
within this gauge. Such transformations (named discrete
transformations in \cite{Salcedo:1998sv}) are of the form
$U(x_0)=\exp(x_0\Lambda)$, where $\Lambda$ is a constant diagonal
matrix. Its eigenvalues $\lambda_j$, $j=1,\ldots,N$ (we consider a
general SU($N$) group in this discussion) are quantized by the
requirement of periodicity in $x_0$.  For quarks, $U(x_0)$ must be
strictly periodic and hence $\lambda_j= 2\pi in_j/\beta$, $n_j\in
\mathbb{Z}$ (the integers $n_j$ are $\bfx$-independent by
continuity). Since under a discrete transformation $A_0(\bfx)\to
A_0(\bfx)+\Lambda$, the eigenvalues of $\log(\Omega)/(2\pi i)$ change
as $\nu_j\to\nu_j-n_j$. In SU(2) this implies that the effective
action in the quark sector must be periodic in $\phi$ with period
$4\pi T$. In the gluon sector, periodicity of $A_\mu(x)$ in $x_0$ only
requires that $U(x_0+\beta)=e^{2\pi ik/N}U(x_0)$, $k=1,\ldots,N$ and
there is an additional symmetry associated to the center of the gauge
group \cite{'tHooft:1979uj,Gross:1981br,Svetitsky:1986ye}. That is,
$\lambda_j= 2\pi i(n_j+k/N)/\beta$ in the absence of quarks (note that
$k$ is both $\bfx$-independent and $j$-independent). The eigenvalues
of $\log(\Om)/(2\pi i)$ change as $\nu_{j\ell}:= \nu_j-\nu_\ell \to
\nu_{j\ell}-n_j+n_\ell$ and the effective action in the gluon sector
must be invariant under such replacement. In SU(2) it corresponds to
periodicity in $\phi$ with period $2\pi T$. From this discussion it
follows that an expansion in powers of $\log(\Omega)$ breaks gauge
invariance under discrete gauge transformations. The local operators
$\vec E_{i\parallel}^2$, $\vec B_{i\parallel}^2$, $\vec E_{i\perp}^2$,
and $\vec B_{i\perp}^2$ are directly gauge invariant.

We can compare these results with those in \cite{Diakonov:2003yy}.
That work goes beyond ours in that we compute the lowest terms in an
expansion in $\widehat D_0$ whereas in \cite{Diakonov:2003yy} all
orders in $A_0$ are retained in the electric sector. On the other
hand, unlike \cite{Diakonov:2003yy}, we treat groups other than SU(2),
our gauge field configurations are not stationary, we consider higher
order terms in the spatial covariant derivatives and we include the
quark sector.

Let us restrict ourselves to stationary gauge configurations and the
gluon sector in SU(2), as in \cite{Diakonov:2003yy}. In a notation
close to that in \cite{Diakonov:2003yy}, the terms of the effective
Lagrangian which are quadratic in $F_{\mu\nu}$, but of any order in
$A_0$, are of the form
\begin{equation}
-f_3(\phi) \vec E^2_{i\parallel}
-f_1(\phi) \vec E^2_{i\perp}
-h_3(\phi) \vec B^2_{i\parallel}
-h_1(\phi) \vec B^2_{i\perp} \,.
\label{eq:3.41}
\end{equation}
To obtain these SU(2) group structure functions in our expansion we
would need to retain terms with two or four spatial indices but any
number of commutators $[A_0,~]$. Nevertheless, in the parallel space
our calculation is complete since all terms of the form $(\widehat
D_0^n F_{\mu\nu})_\parallel$, $n\ge 1$, vanish identically in the
stationary case. This implies that $f_3(\phi)$ and $h_3(\phi)$ do not
get any further contribution beyond those in $\cL_{2,g}(x)$, and
indeed, after passing to the Pauli-Villars scheme with
$\Lambda_{\text{PV}}= e^{1/22}\Lambda_{\overline{\text{MS}}}$, one
verifies that $f_3$ and $h_3$ of \cite{Diakonov:2003yy} are
reproduced.\footnote{After correcting an inaccuracy in the
Pauli-Villars treatment of $f_3$ in \cite{Diakonov:2003yy}. The
correct PV result contains $2\log(\Lambda_{\text{PV}}/4\pi T)-2/11$
instead of just $2\log(\Lambda_{\text{PV}}/4\pi T)$.} $f_1$ is not
reproduced to mass dimension 6, but $h_1$ is reproduced when we retain
mass dimension 4 terms only, since in the magnetic sector the
calculation in \cite{Diakonov:2003yy} introduces ad hoc simplifying
approximations which in practice are equivalent to using
$\cL_{2,g}(x)$.

An important point is that of the periodicity of the structure
functions, also emphasized in \cite{Diakonov:2003yy}. In our
calculation, the coefficients of the local operators will always be
periodic in $\phi$ due to gauge invariance. Yet, this does not imply
that the structure functions themselves should be periodic. The ones
in the parallel sector, which coincide to all orders with the
coefficients in $\cL_{2,g}(x)$, will certainly be periodic, but $f_1$
and $h_1$ will not be periodic in $\phi$. For instance, $h_1$ receives
a contribution from $\cL_{3,g}(x)$ of the form $f(\phi)\vec B_{0i}^2$
(see Appendix \ref{app:C}). The function $f(\phi)$ is periodic and so
this contribution is fully gauge invariant. However, the operator
$f(\phi)\vec B_{0i}^2$ has still to be brought to the standard form in
the (\ref{eq:3.41}). Using $\vec B_{0i}= \vec A_0\times \vec
B_{i\perp}$, it follows that $h_1$ picks up a gauge invariant but non
periodic contribution $\phi^2f(\phi)$. (At this point we disagree with
\cite{Diakonov:2003yy} which notes that $f_1$ needs not be periodic
but requires periodicity of $h_1$.)  We also note that in our
calculation, $f_1$ and $h_1$ are both infrared divergent, whereas in
the calculation of \cite{Diakonov:2003yy} only $h_1$ is
divergent. This should indicate that a resummation to all orders in
$\D_0$ of our expansion may remove spurious infrared divergences.

For SU(3) we present explicit results for the effective Lagrangian up
to mass dimension 4 included. We use the convention
\begin{equation}
A_0= -\frac{i}{2}\lambda_s A_0^s= -\frac{i}{2}\vec\lambda\cdot\vec A_0\,,\quad
F_{\mu\nu}= -\frac{i}{2}\lambda_s F_{\mu\nu}^s\,,\quad \text{etc}
\end{equation}
where $\lambda_s$, $s=1,\dots,8$, are the Gell-Mann matrices.
In the Polyakov gauge
\begin{equation}
   A_0 = -i \frac{\lambda_3}{2}\phi_3-i\frac{\sqrt{3}}{2}\lambda_8 \phi_8 \,.
\end{equation}
The effective Lagrangian from the quark sector can be expressed in
terms of the quantities
\begin{equation}
   \nu_1 = \frac{1}{4 \pi T}(\phi_3+\phi_8)
\,,\quad
   \nu_2 = \frac{1}{4\pi T}(-\phi_3+\phi_8)
\,,\quad
   \nu_3 = -\frac{1}{2\pi T}\phi_8
\end{equation}
as
\begin{eqnarray}
\cL_{0,q} &=& -\frac{\pi^2 T^4 N_f}{12}
\left(-\frac{8}{5}+(1-4\bnu_1^2)^2+(1-4\overline
\nu_2^2)^2+(1-4\bnu_3^2)^2\right) \,,
\end{eqnarray}
and
\begin{eqnarray}
\cL_{2,q} &=& 
\frac{N_f}{24\pi^2}\left(\log\left(\frac{\mu}{4 \pi T}\right) 
- \frac{1}{2}\right) \vec E_i^2 
+ \frac{N_f}{24\pi^2}\log\left(\frac{\mu}{4 \pi T}\right) \vec B_i^2 
\nonumber \\
&& - \frac{N_f}{12 (4\pi)^2}\left(f^-(\nu_1)+f^-(\nu_2)\right)
\left((F_{\mu\nu}^1)^2+(F_{\mu\nu}^2)^2+(F_{\mu\nu}^3)^2 \right)
\nonumber \\
&& - \frac{N_f}{12 (4\pi)^2}\left(f^-(\nu_1)+f^-(\nu_3)\right)
\left((F_{\mu\nu}^4)^2+(F_{\mu\nu}^5)^2 \right)
\nonumber \\
&& - \frac{N_f}{12 (4\pi)^2}\left(f^-(\nu_2)+f^-(\nu_3)\right)
\left((F_{\mu\nu}^6)^2+(F_{\mu\nu}^7)^2 \right)
\nonumber \\
&& - \frac{N_f}{36 (4\pi)^2}\left(f^-(\nu_1)+f^-(\nu_2)+4 f^-(\nu_3)\right)
(F_{\mu\nu}^8)^2
\nonumber \\
&& - \frac{N_f}{6\sqrt{3}(4\pi)^2}\left(f^-(\nu_1)-f^-(\nu_2)\right)
F_{\mu\nu}^3 F_{\mu\nu}^8 \,,
\label{eq:45a}
\end{eqnarray}
where we have defined
\begin{eqnarray}
   f^-(\nu) &=& \psi(\half+\bnu)+\psi(\half-\bnu)
\,,\qquad
   \bnu = (\nu+\frac{1}{2}) \;(\mod 1) -\frac{1}{2}
\,.
\end{eqnarray}

In the gluon sector, we introduce the invariants
\begin{eqnarray}
   \nu_{12} &=& \frac{1}{2 \pi T}\phi_3 \,,
\quad
   \nu_{31} = -\frac{1}{4\pi T}(\phi_3+3\phi_8) \,,
\quad
   \nu_{23} = \frac{1}{4\pi T}(-\phi_3+3\phi_8) \,,
\end{eqnarray}
in terms of which the effective Lagrangian is
\begin{eqnarray}
\cL_{0,g}(x) &=& \frac{4}{3}\pi^2 T^4 \left( -\frac{2}{15} + 
\hnu_{12}^2(1-\hnu_{12})^2 
+ \hnu_{31}^2(1-\hnu_{31})^2 
+ \hnu_{23}^2(1-\hnu_{23})^2 
 \right) \,,
\end{eqnarray}
and
\begin{eqnarray}
\cL_{2,g}(x) &=& 
-\frac{1}{(4\pi)^2}\left(11\log\left(\frac{\mu}{4 \pi T}\right) 
-\frac{1}{2}\right) \vec E_i^2 
-\frac{1}{(4\pi)^2}\left(11\log\left(\frac{\mu}{4 \pi T}\right) 
+\frac{1}{2}\right) \vec B_i^2 
-\frac{T}{4\pi m}\left(\vec E_{i \perp}^2
+\frac{11}{12}\vec B_{i \perp}^2\right)\nonumber \\
&& +\frac{1}{(4\pi)^2}\frac{11}{12}
\left(f^+(0)+f^+(\nu_{12})+\frac{1}{2}f^+(\nu_{31})
+\frac{1}{2}f^+(\nu_{23}) \right)
\left( (F_{\mu\nu}^1)^2+(F_{\mu\nu}^2)^2 \right)
\nonumber \\
&& + \frac{1}{(4\pi)^2}\frac{11}{12}
\left(f^+(0)+\frac{1}{2}f^+(\nu_{12})+f^+(\nu_{31})
+\frac{1}{2}f^+(\nu_{23}) \right)
\left( (F_{\mu\nu}^4)^2+(F_{\mu\nu}^5)^2 \right)
\nonumber \\
&& + \frac{1}{(4\pi)^2} \frac{11}{12}
\left(f^+(0)+\frac{1}{2}f^+(\nu_{12})
+\frac{1}{2}f^+(\nu_{31})+f^+(\nu_{23}) \right)
\left( (F_{\mu\nu}^6)^2+(F_{\mu\nu}^7)^2 \right)
\nonumber \\
&& + \frac{1}{(4\pi)^2} \frac{11}{12}
\left(2 f^+(\nu_{12})+\frac{1}{2}f^+(\nu_{31})
+\frac{1}{2}f^+(\nu_{23}) \right) 
(F_{\mu\nu}^3)^2
\nonumber \\
&& + \frac{1}{(4\pi)^2} \frac{11}{8}
\left(f^+(\nu_{31})+f^+(\nu_{23}) \right)
(F_{\mu\nu}^8)^2
\nonumber \\
&& + \frac{1}{(4\pi)^2}\frac{11}{4\sqrt{3}} 
\left(f^+(\nu_{31})-f^+(\nu_{23})\right)
 F_{\mu\nu}^3 F_{\mu\nu}^8 \,,
\end{eqnarray}
with
\begin{eqnarray}
   f^+(\nu) &=& \psi(\hnu)+\psi(1-\hnu) \, 
\quad (\nu\not\in\mathbb{Z}) \,, 
\quad    \hnu = \nu \; (\mod 1) \,,
\nonumber \\
   f^+(0) &=& - 2\gamma_E \,. 
\end{eqnarray}

Finally, the renormalized tree level is
\begin{equation}
\cL_{\text{tree}}(x) = \frac{1}{4\g^2(\mu)}\,\vec F_{\mu\nu}^2 \,.
\end{equation}

In the stationary case, the most general structure compatible with
SU(3) symmetry, constructed with two $E_i$'s and any number of
$A_0$'s, contains six structure functions (see Appendix \ref{app:D})
\begin{eqnarray}
&&
f_{12}(\phi_3,\phi_8)\left( (E_i^1)^2+(E_i^2)^2\right)
+
f_{45}(\phi_3,\phi_8)\left( (E_i^4)^2+(E_i^5)^2\right)
+
f_{67}(\phi_3,\phi_8)\left( (E_i^6)^2+(E_i^7)^2\right)
\nonumber \\ 
&& \quad
+
f_{33}(\phi_3,\phi_8)\left( E_i^3\right)^2
+
f_{88}(\phi_3,\phi_8)\left( E_i^8 \right)^2
+
f_{38}(\phi_3,\phi_8)\left( E_i^3 E_i^8\right)
\,.
\label{eq:52}
\end{eqnarray}
(And similarly for $B_iB_i$, etc.) Our results for $\cL_2$ are of
this form. Our expressions corresponding to $f_{33}$, $f_{88}$ and
$f_{38}$ are already correct to all orders in $A_0$, since all
$\widehat D_0$ operators cancel in the directions 3 and 8 of the
adjoint space. More generally for any SU($N$) and any structure
function, $A_0$ decomposes $F_{\mu\nu}$ into a parallel component
(that commutes with $A_0$) and a perpendicular component (fully
off-diagonal in the gauge in which $A_0$ is diagonal). The structure
functions not involving perpendicular components depend periodically
on $A_0$ and can be computed exactly using the appropriate finite
order of our expansion (that is, the lowest order at which the
corresponding local operator appears in $\cL$).

In Appendix \ref{app:D} we give further details on the calculation for
SU(3) and SU($N$).

\section{The dimensionally reduced effective theory}
\label{sec:4}

As is well-known, in the high temperature limit non stationary
fluctuations become heavy and are therefore suppressed, and one
expects QCD to behave as an effective three-dimensional theory for the
stationary configurations only
\cite{Ginsparg:1980ef,Appelquist:1981vg,%
Nadkarni:1983kb,Landsman:1989be,Braaten:1996jr,Kajantie:1996dw,%
Shaposhnikov:1996th,KorthalsAltes:2003sv}. Our previous calculation of
the effective action was obtained by separating background from
fluctuation and integrating the latter to one-loop. Clearly, we can
adapt that procedure to obtain the action of the dimensionally reduced
effective theory, to be denoted $\cL^\prime(\bfx)$, by i) using
stationary backgrounds and ii) taking purely non-stationary
fluctuations only, that is, removing the static Matsubara mode in all
frequency summations. In addition, there is a further factor $\beta$
in $\cL^\prime(\bfx)$ from the time integration. Note that ${\mathcal
L}^\prime(\bfx)$ is not the effective action (or Lagrangian) of the
dimensionally reduced theory but its true action (within the one-loop
approximation), in the sense that functional integration over the
stationary configurations with $\cL^\prime(\bfx)$ yields the partition
function. Besides taking $A_\mu$ stationary, we will assume that $A_0$
is small (in particular $|\nu|<1$), which is correct in the high
temperature regime. We will come back to this point later.

The static Matsubara mode is not present in the quark sector, so for
that sector we simply find $\cL^\prime_q(\bfx)=
\beta\cL_q(\bfx)$. Likewise, the removal of the static mode
is irrelevant in the ultraviolet region, hence ${\mathcal
L}^\prime_{\text{tree}}(\bfx)= \beta\cL_{\text{tree}}(\bfx)$
for the renormalized tree level.

As discussed in Appendix \ref{app:B}, the removal of the static mode
in the one-loop gluon sector (and for $|\nu|<1$) corresponds to
replacing $\zeta(1+2\ell+2\epsilon,\nu) \to
\zeta(1+2\ell+2\epsilon,1+\nu)$ in (\ref{eq:50b}). For the effective
potential this means $B_4(\hnu) \to B_4(1+\hnu)$ in (\ref{eq:65}), and
so (dropping an $A_0$-independent term)
\begin{equation}
\cL^\prime_{0,g}(\bfx) =
 \frac{2\pi^2}{3} T^3 \,\htr\left[ \hnu^2 
(1+\hnu^2) \right]
\,,\quad \hnu= \log(\Om)/(2\pi i)\,,\quad
-1 \le \hnu \le 1\,.
\label{eq:4.1}
\end{equation}

The analogous replacement in the mass dimension four and six terms
gives (using the identity $\psi(1+\hnu)+\psi(1-\hnu) =
\psi(\hnu)+\psi(-\hnu) $ )
\begin{equation}
\cL^\prime_{2,g}(\bfx)=
\frac{1}{(4\pi)^2T} \htr \Big[
\frac{11}{12} \left( 2\log(\mu/4\pi T) +\frac{1}{11}
-\psi(\hnu)-\psi(-\hnu) \right) \F_{\mu\nu}^2
-\frac{1}{3} \E_i^2 
\Big]\,,
\label{eq:69a}
\end{equation}
\begin{eqnarray}
\cL^\prime_{3,g}(\bfx) &=&  
\frac{1}{2}\frac{1}{(4\pi)^4 } \frac{1}{T^3}
\htr \bigg[
\Big(\psi^{\prime\prime}(\hnu) 
+ \psi^{\prime\prime}(-\hnu) \Big)
  \\
&& \times
\left( 
\frac{61}{45}\F_{\mu\nu}\F_{\nu\lambda}\F_{\lambda\mu}
+\frac{1}{3}\F_{\lambda\mu\nu}^2
-\frac{1}{30}\F_{\mu\mu\nu}^2
+\frac{3}{5}\F_{0\mu\nu}^2
%+\frac{6}{5}\E_{0i}^2
-\frac{1}{15}\E_{ii}^2
%+\frac{3}{5}\F_{0ij}^2
+\frac{2}{15}\E_i\F_{ij}\E_j
\right)
\bigg]
\,.
\nonumber
\end{eqnarray}
In these expressions $\D_0$ stands for $[A_0,~]$. Note that, having
removed the static mode, $\cL^\prime(\bfx)$ is free from infrared
divergences.

At high temperature the effective potential suppresses configurations
with $\Omega(\bfx)$ far from unity, so by means of a suitable gauge
transformation we can assume that $A_0(\bfx)$ is small \footnote{To
bring $A_0$ to the $|\nu|<1$ basin it will be necessary to use a
discrete gauge transformation, as described in the paragraph after
(\ref{eq:3.40}). Because such transformations are global
($\bfx$-independent) this will be only possible if the original
$A_0(\bfx)$ lies in the same basin (i.e., near the same integer $\nu$)
for all $\bfx$. We assume this, since otherwise $\Omega(\bfx)$ would
be far from unity in the crossover region, thereby increasing the
energy \cite{Bhattacharya:1992qb}.}. In the absence of quarks, the
situation is similar although in this case $\Omega(\bfx)$ lies near a
center of the group element; the center symmetry is spontaneously
broken signaling the deconfining phase
\cite{Polyakov:1978vu,Svetitsky:1986ye,KorthalsAltes:2003sv}. After a
suitable generalized (many-valued) gauge transformation the
configuration can be brought to the small $A_0(\bfx)$ region. It can
be noted that only when $A_0$ is small ($|\nu|<1$) the non static
fluctuations are the heavy ones. If we were to choose the gauge so
that $\nu$ is near some other integer value $n$, the light mode would
be the $n$-th Matsubara mode and integrating out this light mode would
yield a non-local (and so non useful) action for the effective theory.

Because $A_0$ is small, it is standard to expand the
$\cL^\prime(\bfx)$ in powers of $A_0$, using the relation
$\nu=-A_0/(2\pi iT)$ either in the fundamental or the adjoint
representations. We expand up to and including terms of dimension six,
where now $A_0$ coming from $\Omega$ counts as dimension one. Note
that this new counting is free from any ambiguity (although it is in
conflict with explicit gauge invariance).

The effective potential is already a polynomial in $A_0$. From 
(\ref{eq:3.11}) and (\ref{eq:4.1}), we obtain
\begin{equation}
\cL^\prime_0(\bfx)= -\left(\frac{N}{3}+\frac{N_f}{6}\right) T 
\langle A_0^2 \rangle
+\frac{1}{4\pi^2T}\langle A_0^2\rangle^2
+\frac{1}{12\pi^2T}(N-N_f)\langle A_0^4 \rangle \,.
\end{equation}
We have introduced the short-hand notation $\langle X\rangle :=
\tr(X)$ (trace in the fundamental representation) and used the SU($N$)
identity (\ref{eq:4.5}). This result agrees with
\cite{Landsman:1989be,Chapman:1994vk} (there written in the adjoint
representation).

In particular for SU(2) and SU(3), using the identity (\ref{eq:4.5a})
valid for those groups, we find
\begin{equation}
\cL^\prime_0(\bfx)= 
-\left(\frac{N}{3}+\frac{N_f}{6}\right) T \langle A_0^2 \rangle
+\frac{1}{24\pi^2T}(6+N-N_f) \langle A_0^2\rangle^2 \,, \quad N=2,3 \,,
\end{equation}
which reproduces the result quoted in \cite{Shaposhnikov:1996th} and
\cite{KorthalsAltes:2003sv} for $N=3$. We note that consistency
requires to include up to two-loop contributions in the effective
potential \cite{Kajantie:1997tt}.

The terms of dimension four with derivatives come from
$\cL^\prime_2(\bfx)$, given essentially in (\ref{eq:3.31}) (with
$\psi(1-\hnu)\to \psi(-\hnu)$ and an extra factor $\beta$), and
setting $\bnu$ and $\hnu$ to zero. The result can be written as (the
subindex 4 indicates operators of dimension four, all gluon fields
count as mass dimension one)
\begin{equation}
\cL_{(4)}^\prime(\bfx)= 
-\frac{1}{T\g_E^2(T)}\langle E_i^2 \rangle
-\frac{1}{T\g_M^2(T)}\langle B_i^2 \rangle \,.
\end{equation}
(Once again in the fundamental representation.) For the
(chromo)electric and magnetic effective couplings we find
\begin{eqnarray}
\frac{1}{\g_E^2(T)} &=& \frac{1}{\g^2(\mu)}
-2\beta_0(\log(\mu/4\pi T)+\gamma_E)
+\frac{1}{3(4\pi)^2}\left(N+8N_f\left(\log 2-\frac{1}{4}\right)\right)
\nonumber \\
\frac{1}{\g_M^2(T)} &=& \frac{1}{\g^2(\mu)}
-2\beta_0(\log(\mu/4\pi T)+\gamma_E) 
+\frac{1}{3(4\pi)^2}\left(-N+8N_f\log 2\right)
\end{eqnarray}
It is possible to rescale $A_i$ and $A_0$ (with different
renormalization factors) so that $\cL_{(4)}^\prime(\bfx)$ looks like the
zero temperature renormalized tree level (\ref{eq:3.29})
\cite{Nadkarni:1983kb,Landsman:1989be,Chapman:1994vk}. However, we will
work with the original variables.

The result for $\g_M^2(T)$ coincides with \cite{Shaposhnikov:1996th}
for $N=3$. It also agrees with \cite{Chapman:1994vk} (setting
$N_f=0$) assuming a suitable $N$-dependent factor between the scales
$\Lambda$ there and $\mu$ here. The scale independent ratio
\begin{equation}
\frac{\g_E^2(T)}{\g_M^2(T)}=
1-\frac{2}{3}\frac{\g^2(\mu)}{(4\pi)^2}(N-N_f)+O(\g^4) \,,
\end{equation}
found here differs from that reference. On the other hand, 
in analogy with
\begin{eqnarray}
\frac{1}{\g^2(\mu)}  &=& 2\beta_0 \log(\mu/\Lambda_{\overline{\text{MS}}}) \,,
\end{eqnarray}
magnetic and electric thermal $\Lambda$ parameters can be
introduced \cite{Huang:1995cu}
\begin{eqnarray}
\frac{1}{\g_{E,M}^2(T)}  &=& 2\beta_0 \log(T/\Lambda^T_{E,M}) \,,
\end{eqnarray}
which set the scale of high temperatures for both coupling
constants. For the magnetic sector we find
\begin{equation}
\log(\Lambda^T_M/\Lambda_{\overline{\text{MS}}}) =
\gamma_E -\log(4\pi) + \frac{N-8N_f\log 2}{22N-4N_f} \,,
\end{equation}
in agreement with \cite{Huang:1995cu}.

Next, we consider terms of dimension six. They come from
$\cL^\prime_2(\bfx)$ expanding the digamma functions to second order
in $\nu$ and from $\cL^\prime_3(\bfx)$ to zeroth order. From the quark
sector we obtain
\begin{equation}
\cL_{(6),q}^\prime(\bfx) =
\frac{28}{45}\zeta(3)\frac{\beta^3}{(4\pi)^4}N_f \left\langle
F_{\mu\nu}F_{\nu\lambda}F_{\lambda\mu}
+ 6 F_{\mu\mu\nu}^2
+ \frac{9}{2}F_{0\mu\nu}^2
+ 30 A_0^2 F_{\mu\nu}^2
- 3 E_{ii}^2
+ 6 E_iF_{ij}E_j
\right\rangle \,,
\end{equation}
where we have made use of the identity $F_{\lambda\mu\nu}^2=
2F_{\mu\mu\nu}^2-4F_{\mu\nu}F_{\nu\lambda}F_{\lambda\mu}$, valid
inside the functional trace \cite{Chapman:1994vk}. For gluons we have
instead
\begin{equation}
\cL_{(6),g}^\prime(\bfx) =
-\frac{2}{45}\zeta(3)\frac{\beta^3}{(4\pi)^4}\htr\left(
\F_{\mu\nu}\F_{\nu\lambda}\F_{\lambda\mu}
+ \frac{57}{2} \F_{\mu\mu\nu}^2
+ 27\F_{0\mu\nu}^2
+ 165 \hA_0^2 \F_{\mu\nu}^2
- 3 \E_{ii}^2
+ 6 \E_i\F_{ij}\E_j
\right) \,.
\end{equation}
Using (\ref{eq:3.28}) and (\ref{eq:4.5}), this gives for the full
result
\begin{eqnarray}
\cL_{(6)}^\prime(\bfx) &=&
-\frac{2}{15}\frac{\zeta(3)}{(4\pi)^4T^3}\Bigg[
\big( \frac{2}{3}N - \frac{14}{3}N_f \big)
 \langle F_{\mu\nu}F_{\nu\lambda}F_{\lambda\mu} \rangle
+ ( 19 N - 28 N_f )  \langle F_{\mu\mu\nu}^2 \rangle
+ ( 18 N - 21 N_f ) \langle F_{0\mu\nu}^2 \rangle
\nonumber \\ &&
+ ( 110 N - 140 N_f ) \langle A_0^2 F_{\mu\nu}^2 \rangle
- (   2 N - 14 N_f ) \langle E_{ii}^2 \rangle
+ (   4 N - 28 N_f ) \langle E_iF_{ij}E_j \rangle
\nonumber \\ &&
+ 110 \langle A_0^2 \rangle \langle F_{\mu\nu}^2 \rangle
+ 220 \langle A_0 F_{\mu\nu}\rangle^2
\Bigg]
\,.
\end{eqnarray}

For SU(2) and SU(3) the term with $\langle A_0^2 F_{\mu\nu}^2 \rangle$
can be eliminated by using the identity (\ref{eq:4.5a}). In addition,
in SU(2) the term with $\langle F_{0\mu\nu}^2 \rangle$ can also be
removed using (\ref{eq:4.5b}). This produces
\begin{eqnarray}
\cL_{(6)}^\prime(\bfx) &=&
-\frac{2}{15}\frac{\zeta(3)}{(4\pi)^4T^3}\Bigg[
( 3 - 7 N_f ) \left\langle
\frac{2}{3} F_{\mu\nu}F_{\nu\lambda}F_{\lambda\mu}
-\frac{1}{3} F_{0\mu\nu}^2
- 2 E_{ii}^2
+ 4 E_iF_{ij}E_j
\right\rangle
+ ( 57 - 28 N_f )  \langle F_{\mu\mu\nu}^2 \rangle
\nonumber \\ &&
+ \left( 165 - \frac{70}{3} N_f \right)  \left(
\langle A_0^2 \rangle \langle F_{\mu\nu}^2 \rangle
+ 2  \langle A_0 F_{\mu\nu} \rangle^2
\right)
\Bigg]
\,,\qquad \text{for~} N=3
\,,
\end{eqnarray}
\begin{eqnarray}
\cL_{(6)}^\prime(\bfx) &=&
-\frac{4}{15}\frac{\zeta(3)}{(4\pi)^4T^3}\Bigg[
( 2 - 7 N_f ) \left\langle
\frac{1}{3} F_{\mu\nu}F_{\nu\lambda}F_{\lambda\mu}
- E_{ii}^2
+ 2 E_iF_{ij}E_j
\right\rangle
+ ( 19 - 14 N_f ) \langle F_{\mu\mu\nu}^2 \rangle
\nonumber \\ &&
+ ( 74 - 14 N_f ) \langle A_0^2 \rangle \langle F_{\mu\nu}^2 \rangle
+ ( 146 - 21 N_f ) \langle A_0 F_{\mu\nu} \rangle^2
\Bigg]
\,,\qquad \text{for~} N=2
\,.
\end{eqnarray}

$\cL_{(6)}^\prime(\bfx)$ has been computed previously in
\cite{Chapman:1994vk} for the gluon sector and arbitrary number of
colors. Our result agrees with that calculation (and disagrees with
\cite{Wirstam:2001ka}). The dimension six Lagrangian in the quark
sector has been computed in \cite{Wirstam:2001ka} for SU(3), in the
absence of chromomagnetic field ($A_i=0$) and neglecting terms with
more than two spatial derivatives (i.e., neglecting $E_{ii}^2$). Our
result reproduces that calculation in that limit as well.

\section{Conclusions}

In the present work we have developed in full detail the heat kernel
expansion at finite temperature introduced in \cite{Megias:2002vr}. We
have paid special attention to the role played by the untraced
Polyakov loop or thermal Wilson line in maintaining manifest gauge
invariance. This is a highly non trivial problem since preserving
gauge invariance at finite temperature requires infinite orders in
perturbation theory. The conflict between finite order perturbation
theory and finite temperature gauge invariance has been previously
illustrated e.g. in the radiatively induced Chern-Simons action of
$(2+1)$-dimensional fermionic theories \cite{Salcedo:2002pr}.  In the
case where the heat bath is chosen to be at rest the Polyakov loop is
generated by the imaginary time component of the gauge field and can
be regarded as a non-Abelian generalization of the well-known chemical
potential. Actually, we have provided arguments supporting this
interpretation; if the Polyakov loop was absent or represented in
perturbation theory the particle number could not be fixed, as one
expects from standard thermodynamics requirements. The new ingredient
of our technique is that a certain combination of the Polyakov loop
and the temperature has to be treated as an independent variable, in
order to guarantee manifest gauge invariance. This can be done without
fixing the gauge.

An immediate application of our method can be found in QCD at finite
temperature in the region of phenomenological interest corresponding
to the quark-gluon plasma phase. In fact, the heat kernel expansion
corresponds in this case to a high temperature derivative expansion
organized in a very efficient way. In the case of QCD the finite
temperature heat kernel expansion can be applied to compute the
one-loop effective action stemming from the fermion determinant and
from the bosonic determinant corresponding to gluonic fluctuations
around a given background field. As a result we have been able to
reproduce previous partial calculations and to extend them up to terms
of order $T^{-2}$ including the Polyakov loop effects, for a general
gauge group SU($N$). As a by product we have computed the action of
the dimensionally reduced effective theory to the same order. Further
we have studied the emerging group structures in the case of two and
three colors.

\begin{acknowledgments}
This work is supported in part by funds provided by the Spanish DGI
and FEDER founds with grant no. BFM2002-03218, Junta de Andaluc\'{\i}a
grant no. FQM-225, and EURIDICE with contract number
HPRN-CT-2002-00311.
\end{acknowledgments}

\appendix

\section{}
\label{app:A}

Let us establish the commutation rules (\ref{eq:33}). It is sufficient
to consider the case $[X,f]$ since $\D_\mu f$ is a particular
case. Because $f$ is a function of $\Omega$, it is also function of
$D_0$ through the relationship $\Omega=e^{-\beta D_0}$. In fact, it is
better to prove the relation for a general $f(D_0)$ (not necessarily
periodic in its argument). No special property of $D_0$ is required,
so the statement is that, for any two operators $X$ and $Y$ and for
any function $f$,
\begin{eqnarray}
[X,f(Y)] &=& -f^\prime(Y) [Y,X]
+\frac{1}{2}f^{\prime\prime}(Y) [Y,[Y,X]]
+\cdots 
\nonumber \\
&=&
\sum_{n=1}^\infty \frac{(-1)^n}{n!}f^{(n)}(Y)\,D_Y^n(X) \,, 
\qquad  D_Y := [Y,~] \,.
\label{eq:33a}
\end{eqnarray}
It is sufficient to prove this identity for functions of the type
$f(Y)=e^{-\lambda Y}$, where $\lambda$ is a c-number, since the general
case is then obtained through Fourier decomposition.  The r.h.s. of
(\ref{eq:33a}) is
\begin{eqnarray}
\sum_{n=1}^\infty \frac{\lambda^n}{n!}e^{-\lambda Y}D_Y^n(X) 
=
e^{-\lambda Y} (e^{\lambda D_Y}-1)X
=
e^{-\lambda Y} (e^{\lambda Y} X e^{-\lambda Y} -X )
=
[ X ,e^{-\lambda Y}]
\,,
\label{eq:33b}
\end{eqnarray}
which coincides with the l.h.s. of (\ref{eq:33a}). We have used the
well-known identity $e^{D_Y}(X)= e^Y X e^{-Y}$.

\section{}
\label{app:B}

The basic integrals are
\begin{equation}
I^\pm_n(\nu,\alpha):= \int_0^\infty d\tau \tau^{\alpha-1}
\varphi_n^\pm(e^{2\pi i\nu})\,,
\quad \nu,\alpha\in \mathbb{R} \,,\quad
n= 0,1,2,\ldots
\end{equation}
where the functions $\varphi_n$ are defined in (\ref{eq:2.15}) and
$\pm$ refers to the bosonic and fermionic versions, respectively. For
the bosonic version,
\begin{equation}
I^+_n(\nu,\alpha)= \frac{\sqrt{4\pi}}{\beta}\left(\frac{2\pi i}{\beta}\right)^n
\sum_{k\in\mathbb{Z}} (k-\nu)^n
\int_0^\infty d\tau \tau^{\alpha+(n-1)/2 }
e^{-\left(\frac{2\pi}{\beta}\right)^2(k-\nu)^2\tau} \,, 
\quad \nu\not\in\mathbb{Z} \,.
\label{eq:A2}
\end{equation}
We have excluded the case $e^{2\pi i\nu}=1$ which is discussed
below. Integration over $\tau$ gives
\begin{equation}
I^+_n(\nu,\alpha)= 
i^n\frac{\Gamma(\alpha+\frac{n}{2}+\frac{1}{2})}{\Gamma(\frac{1}{2})}
\left(\frac{\beta}{2\pi}\right)^{2\alpha}
\sum_{k\in\mathbb{Z}} \frac{(k-\nu)^n}{|k-\nu|^n}
\frac{1}{|k-\nu|^{2\alpha+1}}
\end{equation}
Defining $\nu=k_0+\widehat\nu$, $0<\widehat\nu<1$, the sum over $k$
can be split into the sum for $k\le k_0$ and another for $k > k_0$. In
terms of the generalized Riemann $\zeta$-function
\cite{Gradshteyn:1980bk} this gives
\begin{equation}
I^+_n(\nu,\alpha)= 
\frac{\Gamma(\alpha+\frac{n}{2}+\frac{1}{2})}{\Gamma(\frac{1}{2})}
\left(\frac{\beta}{2\pi}\right)^{2\alpha}
\big[ (-i)^n \zeta(2\alpha+1,\widehat\nu)
+ i^n \zeta(2\alpha+1,1-\widehat\nu) \big] \,,
\quad 0<\widehat\nu<1 \,, \quad \nu=k_0+\widehat\nu \,,\quad k_0\in\mathbb{Z} \,.
\end{equation}
For the fermionic version, using $\varphi_n^-(\omega)=
\varphi_n^+(-\omega)$ (and so $\nu\to\nu+\frac{1}{2}$), one obtains
\begin{equation}
I^-_n(\nu,\alpha)= 
\frac{\Gamma(\alpha+\frac{n}{2}+\frac{1}{2})}{\Gamma(\frac{1}{2})}
\left(\frac{\beta}{2\pi}\right)^{2\alpha}
\big[ (-i)^n \zeta(2\alpha+1,\half+\bnu)
+ i^n \zeta(2\alpha+1,\half-\bnu) \big] \,,
\quad -\frac{1}{2} < \bnu < \frac{1}{2} \,.
\end{equation}
Note that
\begin{equation}
I^\pm_{2n}(\nu,\alpha)= 
(-1)^n\frac{\Gamma(\alpha+n+\frac{1}{2})}{\Gamma(\alpha+\frac{1}{2})}
I^\pm_0(\nu,\alpha) \,.
\label{eq:B6}
\end{equation}

The formulas are consistent with periodicity and parity
\begin{equation}
I^\pm_n(\nu,\alpha)= I^\pm_n(\nu+1,\alpha)=
(-1)^n I^\pm_n(-\nu,\alpha) \,.
\end{equation}

As discussed in \ref{sec:4}, the dimensionally reduced effective
theory for the stationary configurations, requires to remove the
static mode from the summation over Matsubara frequencies in the
bosonic integrals. This prescription breaks periodicity in $\nu$ but
this is not relevant for the effective theory, since it only describes
the small $A_0$ (or $\nu$) region. (A prescription that preserves
periodicity would be to remove the frequency $k=k_0$ when
$\widehat\nu<\frac{1}{2}$ and $k=k_0+1$ when
$\widehat\nu>\frac{1}{2}$.) The result for the $|\nu|<1$ is
\begin{equation}
I^{\prime +}_n(\nu,\alpha)= 
\frac{\Gamma(\alpha+\frac{n}{2}+\frac{1}{2})}{\Gamma(\frac{1}{2})}
\left(\frac{\beta}{2\pi}\right)^{2\alpha}
\left[
 (-i)^n \zeta(2\alpha+1,1+\nu)
+ i^n \zeta(2\alpha+1,1-\nu) 
\right] \,, 
\qquad -1 < \nu < 1
\,.
\end{equation}

A related issue is that of the infrared divergences for integer
$\nu$. Due to periodicity, we can restrict the discussion to the case
$\nu=0$. For $n\not= 0$, the static Matsubara mode does not contribute
to $I^+_n(\nu,\alpha)$, and so there is no infrared divergence in this
case. On the other hand, in $I^+_0(\nu,\alpha)$, the static mode is
either infrared or ultraviolet divergent. In dimensional
regularization such an integral ($\nu=k=n=0$ in (\ref{eq:A2})) is
defined as zero since it has no natural scale \cite{Collins:1984bk}.
So for all $n$ the result is equivalent to removing the static mode
\begin{equation}
I^+_n(\nu,\alpha)= 
I^{\prime +}_n(0,\alpha)= \Bigg\{ \matrix{
(-1)^{n/2} 2\pi^{-1/2}
\Gamma(\alpha+\frac{n}{2}+\frac{1}{2})
\left(\beta/2\pi\right)^{2\alpha} \zeta(2\alpha+1) \,, &&
\text{even\ } n \cr 0 \,, && \text{odd\ } n } 
\quad 
\text{for\ } \nu\in\mathbb{Z}
\,.
\label{eq:B10}
\end{equation}

Alternatively one can regulate the infrared divergence by adding a
cutoff function $e^{-m^2\tau}$ ($m\to 0$) in the $\tau$ integral. This
amounts to adding a contribution $\sqrt{4\pi}
\Gamma(\alpha+1/2)/(\beta m^{2\alpha+1})$ in $I^+_0(\nu,\alpha)$ for
integer $\nu$.

\section{}
\label{app:C}
In this appendix we present results for SU(2) in both sectors,
including all terms of mass dimension 6. All results are given in the
$\overline{\text{MS}}$ scheme. In these formulas we have allowed for
an explicit infrared cut off $m$, as commented at the end of Appendix
\ref{app:B}. The results with strict dimensional regularization are
recovered by removing all infrared divergent terms from the
formulas. The conventions are those of subsection \ref{subsec:3.d}.

\begin{equation}
\cL_{\text{tree}}(x) = \frac{1}{4\g^2(\mu)}\,\vec F_{\mu\nu}^2 \,,
\end{equation}

\begin{eqnarray}
\cL_{0,g}(x) &=& 
\frac{\pi^2 T^4}{3} 
\left(-\frac{1}{5}+4\hnu^2 (1-\hnu)^2\right)
\,,
\end{eqnarray}

\begin{eqnarray}
 \cL_{2,g}(x) &=& -\frac{11}{96 \pi^2}\left(\frac{1}{11}+2\log\left(\frac{\mu}{4\pi T}\right)-\psi(\hnu)-\psi(1-\hnu)\right) 
\vec F_{\mu\nu \parallel}^2
\nonumber \\
&&- \frac{11}{96 \pi^2}\left(\frac{\pi T}{m}+\frac{1}{11}+2\log\left(\frac{\mu}{4\pi T}\right)+\gamma_E-\frac{1}{2}\psi(\hnu)-\frac{1}{2}\psi(1-\hnu)\right) \vec F_{\mu\nu \perp}^2
\nonumber \\
&&+\frac{1}{24\pi^2}\vec{E}_i^2 -\frac{1}{48\pi^2}\left(\frac{\pi T}{m}\right)\vec{E}_{i \perp}^2 \,,
\end{eqnarray}

\begin{eqnarray}
 \cL_{3,g}(x) &=&\frac{61}{2160\pi^2}\left(\frac{1}{4\pi T}\right)^2\left(8\left(\frac{\pi T}{m}\right)^3+2\zeta(3)-\psi^{\prime\prime}(\widehat\nu)-\psi^{\prime\prime}(1-\widehat\nu)\right)(\vec{F}_{\mu\nu} \times \vec{F}_{\nu\alpha})\cdot \vec{F}_{\alpha\mu} 
\nonumber \\
&&-\frac{1}{48\pi^2}\left(\frac{1}{4\pi T}\right)^2\left(\psi^{\prime\prime}(\widehat\nu)+\psi^{\prime\prime}(1-\widehat\nu)\right)\vec{F}_{\lambda\mu\nu \parallel}^2
\nonumber \\
&&+\frac{1}{96\pi^2}\left(\frac{1}{4\pi T}\right)^2\left(16\left(\frac{\pi T}{m}\right)^3+4\zeta(3)-\psi^{\prime\prime}(\widehat\nu)-\psi^{\prime\prime}(1-\widehat\nu)\right)\vec{F}_{\lambda\mu\nu \perp}^2
\nonumber \\
&&+\frac{1}{480\pi^2}\left(\frac{1}{4\pi T}\right)^2\left(\psi^{\prime\prime}(\widehat\nu)+\psi^{\prime\prime}(1-\widehat\nu)\right)\vec{F}_{\mu\mu\nu \parallel}^2
\nonumber \\
&&-\frac{1}{960\pi^2}\left(\frac{1}{4\pi T}\right)^2\left(16\left(\frac{\pi T}{m}\right)^3+4\zeta(3)-\psi^{\prime\prime}(\widehat\nu)-\psi^{\prime\prime}(1-\widehat\nu)\right)\vec{F}_{\mu\mu\nu \perp}^2
\nonumber \\
&&-\frac{3}{80\pi^2}\left(\frac{1}{4\pi T}\right)^2\left(\psi^{\prime\prime}(\widehat\nu)+\psi^{\prime\prime}(1-\widehat\nu)\right)\vec{F}_{0\mu\nu \parallel}^2\nonumber \\
&&+\frac{3}{160\pi^2}\left(\frac{1}{4\pi T}\right)^2\left(-8\left(\frac{\pi T}{m}\right)^3+4\zeta(3)-\psi^{\prime\prime}(\widehat\nu)-\psi^{\prime\prime}(1-\widehat\nu)\right)\vec{F}_{0\mu\nu \perp}^2
\nonumber \\
&&-\frac{1}{10\pi^2}\left(\frac{1}{4\pi T}\right)^2\left(\frac{\pi T}{m}\right)^3 \vec{E}_{0i \perp}^2
\nonumber \\ 
&&+\frac{1}{240\pi^2}\left(\frac{1}{4\pi T}\right)^2\left(\psi^{\prime\prime}(\widehat\nu)+\psi^{\prime\prime}(1-\widehat\nu)\right)\vec{E}_{ii \parallel}^2\nonumber \\
&&-\frac{1}{480\pi^2}\left(\frac{1}{4\pi T}\right)^2\left(-8\left(\frac{\pi T}{m}\right)^3+4\zeta(3)-\psi^{\prime\prime}(\widehat\nu)-\psi^{\prime\prime}(1-\widehat\nu)\right)\vec{E}_{ii \perp}^2
\nonumber \\
&&+\frac{1}{240\pi^2}\left(\frac{1}{4\pi T}\right)^2\left(\psi^{\prime\prime}(\widehat\nu)+\psi^{\prime\prime}(1-\widehat\nu)\right)\varepsilon_{ijk}(\vec{E}_i\times \vec{E}_j)\cdot \vec{B}_k
\nonumber \\
&&+\frac{1}{240\pi^2}\left(\frac{1}{4\pi T}\right)^2\left(8\left(\frac{\pi T}{m}\right)^3-4\zeta(3)-\psi^{\prime\prime}(\widehat\nu)-\psi^{\prime\prime}(1-\widehat\nu)\right)\varepsilon_{ijk}(\vec{E}_{i \perp}\times\vec{E}_{j \perp})\cdot \vec{B}_{k \parallel}
\,,
\end{eqnarray}

\begin{eqnarray}
\cL_{0,q}(x) &=& \frac{2}{3} \pi^2  T^4 N_f \left(\frac{2}{15}-\frac{1}{4}(1-4 \bnu^2)^2\right)
\,,
\end{eqnarray}

\begin{eqnarray}
 \cL_{2,q}(x) &=&
\frac{N_f}{96 \pi^2}\left(2\log\left(\frac{\mu}{4\pi T}\right)
-\psi(\half+\bnu)-\psi(\half-\bnu)\right) \vec{F}_{\mu\nu}^2 
-\frac{N_f }{48 \pi^2}\vec{E}_i^2
\,,
\end{eqnarray}

\begin{eqnarray}
 \cL_{3,q}(x) &=& \frac{N_f}{960\pi^2}\left(\frac{1}{4\pi T}\right)^2
\big(\psi^{\prime\prime}(\half +\bnu)+\psi^{\prime\prime}(\half-\bnu)\big)
\nonumber \\
&&
\times 
\Big(
\frac{16}{3}(\vec{F}_{\mu\nu} \times \vec{F}_{\nu\alpha}) \cdot \vec F_{\alpha\mu}+\frac{5}{2}\vec{F}_{\lambda\mu\nu}^2
-\vec{F}_{\mu\mu\nu}^2
-2\varepsilon_{ijk}(\vec{E}_i\times\vec{E}_j) \cdot \vec{B}_k
+3\vec{F}_{0\mu\nu}^2
-2\vec{E}_{ii}^2
\Big)
\,.
\end{eqnarray}

It can be noted that the quark terms do not distinguish between
parallel and perpendicular components. This is due to the fact that in
SU(2) an even function of $\bnu$ (or any other element of
su(2)) in the fundamental representation is necessarily a
c-number. Since the $\varphi_n$ functions involved to mass dimension 6
are all even, the $\bnu$ dependence gets out of the trace in
(\ref{eq:14}) and (\ref{eq:16}) and $A_0$ is no longer a privileged
direction in color space. This mechanism does not act in the adjoint
representation, i.e., in the gluon sector, nor for other SU($N$)
groups (cf. (\ref{eq:45a})).

The infrared divergence is tied to $\nu$ integer, so it does not exist
for fermions, and also cancels in all gluon terms involving only
parallel components.

\section{}
\label{app:D}

In SU($N$) the gauge can be chosen so that $A_0$ is diagonal. This
form is unique (up to permutation of eigenvalues) and produces $N-1$
quantities invariant under SU($N$) ($\phi_3$, $\phi_8$ for SU(3)). If
$X$ represents $F_{\mu\nu}$ or any other element of su($N$)
($X^\dagger=-X$, $\tr(X)=0$) with $N^2-1$ independent components, we
can use the remaining gauge freedom (the $N-1$ gauge transformations
which leave $A_0$ diagonal) to fix $N-1$ of these components. This
adds $(N^2-1)-(N-1)$ new invariants involving $X$ (and $A_0$). Of
these, $N-1$ are linear in $X$ (the diagonal components of $X$),
$N(N-1)/2$ are quadratic and $(N-1)(N-2)/2$ are cubic. For instance,
in SU(3), under a diagonal gauge transformation
\begin{equation}
X=\left( \matrix{ x & a & b \cr -a^* & y & c \cr -b^* & -c^* & -x-y }
\right)
\to
\left( \matrix{ x & e^{i(\alpha-\beta)}a & 
 e^{i(2\alpha+\beta)} b \cr - e^{-i(\alpha-\beta)}a^* & y & 
 e^{i(\alpha+2\beta)} c \cr - e^{-i(2\alpha+\beta)}b^* & 
- e^{-i(\alpha+2\beta)}c^* & -x-y }
\right)
\end{equation}
the invariants are $x$, $y$, $aa^*$, $bb^*$, $cc^*$ and $ab^*c$ (the
last one is complex but its modulus is not independent). For $X=E_i$
this gives the six structure functions in (\ref{eq:52}). Each further
vector $Y\in\text{su($N$)}$ produces new $N^2-1$ invariants.

For computing the traces in the adjoint representation one possibility
is to use the adjoint basis $(T^s)_{rt}= f_{rst}$, such that to
$F_{\mu\nu}= F_{\mu\nu}^st_s$ ($t_s= \lambda_s/2i$) in the fundamental
representation it corresponds $\F_{\mu\nu}= F_{\mu\nu}^sT_s$ in the
adjoint one. We have also used an alternative approach, as
follows. The elements of su($N$), such as the gluon quantum
fluctuation $a_\mu$ are $N\times N$ matrices, $(a_\mu)_{a\da}$. From
the action $\F_{\mu\nu}(a_\lambda)=[F_{\mu\nu},a_\lambda]$, it follows
\begin{equation}
(\F_{\mu\nu})_{a\da,b\db}= 
(F_{\mu\nu})_{ab}\delta_{\da\db}-\delta_{ab}(F_{\mu\nu})_{\db\da}
\,,\qquad a,b,\da,\db= 1,\ldots,N\,.
\end{equation}
In matrix notation this can be written as $\F_{\mu\nu}=
F_{\mu\nu}\otimes 1-1\otimes F_{\mu\nu}^T = F_{\mu\nu}\otimes
1+1\otimes F_{\mu\nu}^*$, or even, in shorter form,
\begin{equation}
\F_{\mu\nu}= F_{\mu\nu}- F_{\mu\nu}^T
= F_{\mu\nu}+ F_{\mu\nu}^*
\end{equation}
understanding that $F_{\mu\nu}^T$ or $F_{\mu\nu}^*$ always refer to
the dotted space. Similarly, $\hA_\mu= A_\mu-A_\mu^T= A_\mu+A_\mu^*$.
Since dotted and undotted operators commute, it follows that the
$\Om=\Omega\otimes\Omega^*=\Omega\otimes\Omega^{-1T}$ for the Polyakov
loop. In the Polyakov gauge ($A_0$ stationary and diagonal) $\Omega$
is diagonal $(\Omega)_{ab}=\omega_a\delta_{ab}$ and $\Om$ is also
diagonal in that basis, $(\Om)_{a\da,b\db}= \omega_{a\da}
\delta_{ab}\delta_{\da\db}$, with $\omega_{a\da}=
\omega_a\omega^{-1}_\da$.

From the point of view of the gauge group, the computation of the
trace in the adjoint space involves only four different structures
appearing in $\hb_{0,g}$,$\hb_{2,g}$, and $\hb_{3,g}$. These are
\begin{eqnarray}
\htr(f(\Om)) &=&
{\sum_{a\da}}^\prime f(\omega_{a\da}) \,,
\nonumber \\
\htr(f(\Om)\F_{\mu\nu}^2) &=&
\sum_{a\da}f(\omega_{a\da}) \big[ (F_{\mu\nu}^2)_{aa}
+(F_{\mu\nu}^2)_{\da\da} - 2 (F_{\mu\nu})_{aa}(F_{\mu\nu})_{\da\da}
\big] \,,
\nonumber \\
\htr(f(\Om)\F_{\mu\nu}\F_{\nu\lambda}\F_{\lambda\mu})) &=&
\sum_{a\da}f(\omega_{a\da}) \big[ 
(F_{\mu\nu}F_{\nu\lambda}F_{\lambda\mu})_{aa}
+(F_{\mu\nu}F_{\nu\lambda}F_{\lambda\mu})_{\da\da}
-(F_{\mu\nu})_{aa}(F_{\nu\lambda}F_{\lambda\mu})_{\da\da}
-(F_{\mu\nu}F_{\nu\lambda})_{aa}(F_{\lambda\mu})_{\da\da}
\big] \,,
\nonumber \\
\htr(f(\Om)\E_i\F_{ij}\E_j) &=&
\sum_{a\da}f(\omega_{a\da}) \big[ 
(E_iF_{ij}E_j)_{aa} 
+(E_iF_{ij}E_j)_{\da\da}
-(E_i)_{aa}([F_{ij},E_j])_{\da\da}
-([E_i,F_{ij}])_{aa}(E_j)_{\da\da}
\nonumber \\
&&\qquad \quad 
- (E_iE_j)_{aa}(F_{ij})_{\da\da}
- (F_{ij})_{aa}(E_iE_j)_{\da\da}
\big]
\,.
\end{eqnarray}
($\sum_{a\da}^\prime$ in the first equation indicates that one of the
$N$ modes with $a=\da$ should not be included. This removes the
singlet mode present in U($N$) but not SU($N$). The singlet mode does
not contribute in the other formulas.) Often,
$f(\omega)=f(\omega^{-1})$ (i.e. $f(\omega_{a\da})$ is symmetric in
$a,\da$), but this property has been not used here. It can be observed
that the contributions $a=\da$, which correspond to $\Om=1$ and are
afflicted by infrared divergences, cancel in the subspace parallel
(i.e. for $F_{\mu\nu}$ diagonal in the Polyakov gauge).

Useful SU($N$) identities are ($\langle~\rangle$ stands for trace in the fundamental representation)
\begin{equation}
\htr\left(\widehat{X}^2\right) = 2N\langle X^2\rangle \,,
\quad X\in\text{su($N$)}\,,
\label{eq:3.28}
\end{equation}
\begin{equation}
\htr(\widehat{X}^2\widehat{Y}^2)= 2N\langle X^2Y^2\rangle +
2\langle X^2\rangle\langle Y^2\rangle+4\langle XY\rangle^2 \,,\quad X,Y\in\text{su($N$)}\,,
\label{eq:4.5}
\end{equation}
\begin{equation}
\langle X^2Y^2 \rangle = -\frac{1}{6}\langle [X,Y]^2 \rangle 
+ \frac{1}{6}\langle X^2\rangle\langle Y^2\rangle 
+ \frac{1}{3}\langle XY\rangle^2
\quad X,Y\in\text{su($3$)} 
\,,
\label{eq:4.5a}
\end{equation}
\begin{equation}
\langle [X,Y]^2 \rangle = -2\langle X^2\rangle\langle Y^2\rangle 
+ 2\langle XY\rangle^2
\quad X,Y\in\text{su($2$)} 
\,.
\label{eq:4.5b}
\end{equation}

%\bibliography{Refs}

\end{document}